\documentclass[acmtog]{acmart}

\usepackage{booktabs} 
\usepackage{multirow}

\AtBeginDocument{%
  }

\acmJournal{TOG}
\acmYear{2025} \acmVolume{44} \acmNumber{4} \acmArticle{} \acmMonth{8} \acmPrice{}\acmDOI{10.1145/3730825}

\begin{document}

\title{Facial Appearance Capture at Home with Patch-Level Reflectance Prior}


\author{Yuxuan Han}
\affiliation{%
  \institution{School of Software and BNRist, Tsinghua University}
  \city{Beijing}
  \country{China}
}
\email{hanyx22@mails.tsinghua.edu.cn}
\orcid{0000-0002-2844-5074}

\author{Junfeng Lyu}
\affiliation{%
  \institution{School of Software and BNRist, Tsinghua University}
  \city{Beijing}
  \country{China}
}
\email{lyujunfeng@gmail.com}
\orcid{0009-0007-8083-2948}

\author{Kuan Sheng}
\affiliation{%
 \institution{ShanghaiTech University and Deemos Technology Co., Ltd.}
 \city{Shanghai}
 \country{China}
}
\email{shengkuan@shanghaitech.edu.cn}
\orcid{0009-0005-8065-2475}

\author{Minghao Que}
\affiliation{%
  \institution{School of Software and BNRist, Tsinghua University}
  \city{Beijing}
  \country{China}
}
\email{quemh22@mails.tsinghua.edu.cn}
\orcid{0009-0005-5490-8940}

\author{Qixuan Zhang}
\affiliation{%
 \institution{ShanghaiTech University and Deemos Technology Co., Ltd.}
 \city{Shanghai}
 \country{China}
}
\email{zhangqx1@shanghaitech.edu.cn}
\orcid{0000-0002-4837-7152}

\author{Lan Xu}
\affiliation{%
 \institution{ShanghaiTech University}
 \city{Shanghai}
 \country{China}
}
\email{xulan1@shanghaitech.edu.cn}
\orcid{0000-0002-8807-7787}

\author{Feng Xu}
\affiliation{%
  \institution{School of Software and BNRist, Tsinghua University}
  \city{Beijing}
  \country{China}
}
\email{xufeng2003@gmail.com}
\orcid{0000-0002-0953-1057}

\renewcommand{\shortauthors}{Han et al.}

\begin{abstract}
  Existing facial appearance capture methods can reconstruct plausible facial reflectance from smartphone-recorded videos. However, the reconstruction quality is still far behind the ones based on studio recordings. This paper fills the gap by developing a novel daily-used solution with a co-located smartphone and flashlight video capture setting in a dim room. To enhance the quality, our key observation is to solve facial reflectance maps within the data distribution of studio-scanned ones. Specifically, we first learn a diffusion prior over the Light Stage scans and then steer it to produce the reflectance map that best matches the captured images. We propose to train the diffusion prior at the patch level to improve generalization ability and training stability, as current Light Stage datasets are in ultra-high resolution but limited in data size. Tailored to this prior, we propose a patch-level posterior sampling technique to sample seamless full-resolution reflectance maps from this patch-level diffusion model. Experiments demonstrate our method closes the quality gap between low-cost and studio recordings by a large margin, opening the door for everyday users to clone themselves to the digital world. Our code will be released at \href{https://github.com/yxuhan/DoRA}{https://github.com/yxuhan/DoRA}.
\end{abstract}

\ccsdesc[500]{Computing methodologies~Rendering}

\keywords{Face Modeling, Appearance Capture, Diffusion Model.}


\begin{teaserfigure}
\centering
    \includegraphics[width=1.0\textwidth]{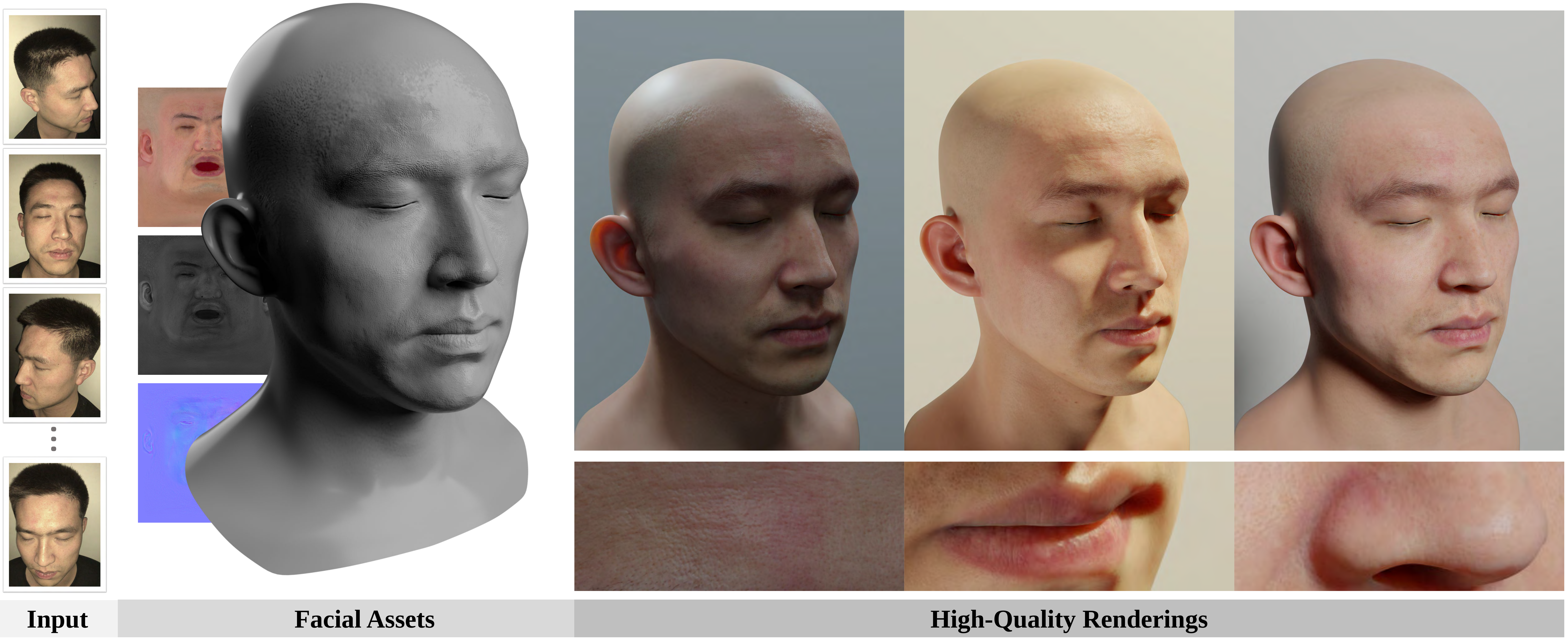}
    \caption{
    We propose a novel method for low-cost high-quality facial appearance capture. 
    Given a single co-located smartphone and flashlight sequence captured in a dim room as the input, our method can reconstruct high-quality facial assets, which can be exported to common graphics engines like Blender for photo-realistic rendering.
    As shown in the figure above, our method can faithfully reconstruct detail patterns on different facial regions like the forehead, lip, and nose.
    }
    \label{fig:teaser}
\end{teaserfigure}

\maketitle

\section{Introduction}
Facial appearance capture methods usually estimate facial reflectance information such as diffuse albedo, specular albedo, and normal, from face observations like images and videos.
As one key technique to clone our human beings to the digital world, it enables various applications in digital movies, game production, Metaverse, and many more.
Despite advances in recent years, this problem remains challenging. 
On the one hand, due to the complex nature of skin~\cite{ma2007rapid,wann2023practical,weyrich2006analysis}, it is extremely difficult to reconstruct high-quality facial reflectance from images.
On the other hand, as we humans are susceptible to facial traits, we can detect even the slightest artifacts on faces, leading to the uncanny valley effect.

To address this problem, some works build specialized apparatus in studios to capture facial appearance under well-controlled setups~\cite{ma2007rapid,ghosh2011multiview,riviere2020single}.
Typically, these methods use high-definition DSLR cameras to capture images and deploy polarization filters for diffuse-specular separation.
Although demonstrating production-level results~\cite{alexander2009digital,alexander2013digital}, they are only viable for a few professional users as on-site data capture is very costly.
Recent works propose to democratize the facial appearance capture process with smartphone recordings~\cite{han2024cora,wang2023sunstage,rainer2023neural}.
Although easy to use, the reconstruction quality is still far behind the studio-based methods.
The reasons are twofold.
On the one hand, the image quality of the smartphone camera is significantly inferior to that of DSLR cameras used by the studio-based method.
On the other hand, without polarization filters for explicit diffuse-specular separation, the inherent entanglement of these two components leads to degraded estimation~\cite{ma2007rapid}.

In this paper, our goal is to fill the gap between the two worlds, \emph{i.e.} developing a method that is easy to use by {everyday users} at home while achieving comparable quality to the high-budget method in the studio.
Following CoRA~\cite{han2024cora}, we take a single co-located smartphone and flashlight sequence captured in a dim room as input.
This setting can be used by everyday users but still records as much information as possible.
Given that the quality gap arises from insufficient observations, \emph{i.e.} low-quality image capture and the absence of polarization filters for separating diffuse and specular components, our key idea is to introduce a high-quality reflectance prior to enhance reconstruction quality.
In brief, we train a diffusion model~\cite{ho2020denoising,Karras2022edm} on Light Stage scans to model the joint distribution of high-quality diffuse albedo, specular albedo, and normal in the UV space.
During reconstruction, we solve the reflectance maps within the distribution modeled by this diffusion prior, where we apply diffusion posterior sampling (DPS)~\cite{chung2022diffusion} to steer the diffusion model to generate a reflectance map that satisfies the photometric loss.

However, this is not easy to achieve.
Firstly, training a diffusion model on the Light Stage dataset is challenging, as the reflectance maps are in ultra-high resolutions but the data size is limited.
To address this problem, we randomly crop patches from the original maps and model the reflectance distribution at the patch level.
Considering that the global information is lost after cropping, we condition the diffusion model on the UV coordinate map of the patches.
In this way, we can train the diffusion model at a significantly lower resolution like 256 by 256 with a large corpus of training data while keeping both the local and global prior knowledge.

Secondly, applying such a prior model in reconstructing high-resolution reflectance maps is also challenging.
In theory, given the diffusion prior trained on reflectance patches, we can directly apply DPS~\cite{chung2022diffusion} to sample the full-resolution map as the UNet-based network can take any resolution input.
However, it is infeasible in practice due to the memory limitation of the graphics card\footnote{We find an 80G NVIDIA A800 can process on no more than 1536 by 1536 resolution while the typical resolution of a production-usage reflectance map is 4K by 4K.}.
To reduce memory costs, we split the full-resolution map into overlapped patches and propose a patch-based DPS technique to sample them collaboratively.
Specifically, we alternate between the DPS~\cite{chung2022diffusion} to compute the updating direction for each patch independently and blend the overlapped regions via Tiled Diffusion~\cite{bar2023multidiffusion,jimenez2023mixture}.
This way, we obtain seamless full-resolution results while sampling at the patch level.

In conclusion, our main contributions include:
\begin{itemize}
    \item A novel facial appearance capture method that closes the quality gap between low-cost and studio recordings by a large margin. 
    \item The first diffusion prior to model the joint distribution of high-quality patch-level diffuse albedo, specular albedo, and normal map with the UV coordinate map as a condition.
    \item A novel memory-efficient posterior sampling technique to solve the full-res reflectance map within the distribution modeled by our patch-level diffusion prior.
\end{itemize}
Experiments demonstrate our method closes the quality gap between low-cost and studio recordings by a large margin, opening the door for everyday users to clone themselves to the digital world with high fidelity.

\section{Related Work}
\subsection{Facial Appearance Capture}
Facial appearance capture has attracted much attention in recent years~\cite{klehm15star}.
The goal is to estimate reflectance parameters such as diffuse albedo, specular albedo, and normal from observed images.
However, this problem is quite challenging.
Due to the translucent nature of the skin, a portion of light would enter the skin and scatter multiple times beneath the skin~\cite{debevec2000acquiring}.
These lights cause the diffuse component. 
Solving reflectance parameters from it leads to blurry results as it is the average effect of neighborhood points.
Another portion of light directly reflects on the skin's surface, which causes the specular component.
We can estimate accurate specular albedo and normal from it, as it is solely associated with the shading point.
Thus, plausible disentanglement of diffuse and specular components is required to ensure high-quality results~\cite{ma2007rapid,ghosh2011multiview}.

Traditional methods build specialized apparatus in the studio to capture data in a well-posed setup where the captured images can be easily disentangled into diffuse and specular.
\citet{weyrich2006analysis} build a Light Stage~\cite{debevec2000acquiring} to capture multi-view One-Light-At-a-Time (OLAT) images of the subject.
They separate diffuse and specular based on the observation that specular reflection vanishes at some view-light direction pairs.
\citet{ma2007rapid} deploy polarization filters in front of the light and camera on the Light Stage.
By controlling the polarization state, they can capture diffuse or specular-only images.
\citet{ghosh2011multiview} further extend this work to support multi-view data capture.
More recently, some works~\cite{riviere2020single,lattas2022practical} design cheaper data capture systems to reduce costs.
Although these methods demonstrate impressive results~\cite{alexander2009digital,alexander2013digital}, they are only viable for a few professional users with high budgets.

Recent methods propose to democratize the facial appearance capture process.
A group of works pretrain a neural network on a Light Stage dataset to learn the inverse rendering process~\cite{han2023learning,lattas2020avatarme,yamaguchi2018high,Dib_2024_CVPR,Paraperas_2023_ICCV}. 
At inference time, they can reconstruct a relightable scan from a single-view facial image in the wild.
Although easy to use, in terms of resembling the captured subject, their results are still far behind the studio-based ones, as they take only a single image as input.
Another group of works~\cite{han2024cora,wang2023sunstage,rainer2023neural,bharadwaj2023flare} reconstructs facial appearance from a smartphone video captured around the face. 
The representative work, CoRA~\cite{han2024cora}, takes a co-located smartphone and flashlight sequence captured in a dim room as input.
Compared to the single-view methods, these works can reconstruct higher fidelity results.
To improve the quality, \citet{azinovic2023high} deploy polarization filters on the smartphone flashlight and camera for better diffuse-specular separation.
Although they demonstrate better results, the requirement of polarization filters makes their system hard to use for everyday users.

In contrast to \citet{azinovic2023high}, we only use a smartphone for data capture as CoRA~\cite{han2024cora} to make our method easy to use. 
Instead, we enhance the diffuse-specular separation in a data-driven fashion, \emph{i.e.} using a high-quality reflectance prior learned from a Light Stage dataset.
Perhaps the most similar work to us is \citet{hifi3dface2021tencentailab}.
They learn a PCA model on high-quality facial scans to model the reflectance prior at full resolution and use it to constrain the reconstruction.
However, limited by the capability of the PCA model, their method cannot reconstruct person-specific facial traits like nevus.
Compared to \citet{hifi3dface2021tencentailab}, we adopt the diffusion model as the prior with a more powerful representation capability.
However, the diffusion model requires significantly more training data than the PCA.
To this end, we learn the diffusion prior at the patch level and develop tailored techniques to generate full-resolution results using this patch-level prior.

\subsection{Diffusion Model for Inverse Problems}
In inverse problems, the goal is to recover the unknown data sample given measurements and a corrupted forward model~\cite{daras2024survey}.
For example, in image super-resolution, the unknown is the high-res image, the measurement is the given low-res image, and the corrupted forward model is the down-sampling operator.
The inverse problems are usually ill-posed, as the given measurements and the corrupted forward model usually cannot derive a unique solution.
Thus, priors are important in inverse problems as they can drive the optimization to a plausible solution.

Recently, the diffusion model~\cite{ho2020denoising,Karras2022edm} has demonstrated remarkable performance in content generation~\cite{po2024state}, including images, videos, and 3D assets.
Motivated by this, researchers propose to use a pretrained diffusion model as a prior in the inverse problem.
ReconFusion~\cite{wu2023reconfusion} uses a diffusion prior in the few-shot view synthesis task.
The goal is to train a NeRF given a few input images.
They pretrain a diffusion model for novel view synthesis on a large-scale dataset.
During reconstruction, they use this model to generate pseudo-views to augment the NeRF training set.
Intrinsic Anything~\cite{chen2025intrinsicanything} adopts a diffusion prior in the inverse rendering problem.
The goal is to reconstruct geometry and reflectance from multi-view images captured around the object under unknown lighting.
They pretrain a diffusion model for intrinsic decomposition and use this model to predict the reflectance map for each image as the pseudo ground truth.
Although impressive results are demonstrated, these works typically require large-scale datasets like Objaverse~\cite{deitke2023objaverse} to cover all possible pairs of data samples and measurements for pretraining.

Another group of works directly learns a diffusion prior over the data sample~\cite{chung2022diffusion,rout2023secondorder,daras2024survey}.
In this way, they can largely reduce the demand for data size.
In addition, once trained, the prior can be used to solve different tasks.
Diffusion Posterior Sampling (DPS)~\cite{chung2022diffusion} uses a diffusion prior for image restoration.
They train an unconditional diffusion model to model the distribution of high-res images.
Given a corrupted image, they steer the diffusion model to generate an image that best matches the measurement after being processed by the corrupted forward model.
To this end, they introduce a guidance term to the sampling process of the diffusion model, which is derived from the discrepancy between the corrupted generated sample and the measurement.
Diffusion Posterior Illumination (DPI)~\cite{lyu2023diffusion} further applies DPS to the inverse rendering task.
They train a diffusion model over HDR environment maps.
During reconstruction, they adopt DPS to sample an environment map within the distribution modeled by the diffusion model that minimizes the photometric loss, and simultaneously optimize the reflectance.

Motivated by DPI~\cite{lyu2023diffusion}, we leverage the diffusion model as a prior in our inverse rendering problem, \emph{i.e.} low-cost facial appearance capture.
In contrast, we focus on improving the quality of the reflectance map while DPI uses the diffusion model to model the ambiguity in lighting estimation.
In addition, our diffusion prior is trained on patches due to the limited data size of the Light Stage dataset, while DPI directly trains on large-scale environment map datasets.
To use this patch-level diffusion prior, we propose a novel sampling technique.
To the best of our knowledge, we are the first to solve the reflectance map within the distribution of a patch-level diffusion model.

\subsection{Face Modeling at Patch-Level}
As the size of high-quality 3D facial datasets is usually limited, previous works propose to model faces at patch level to augment the training data.
AvatarMe~\cite{lattas2020avatarme,lattas2021avatarme++} trains an image-to-image translation network at the patch level to translate the UV diffuse albedo map into other reflectance maps such as the specular albedo. 
After training, they directly apply this network to process the full-resolution map.
It can obtain seamless results as the neural network implicitly fuses the border of patches at test time.
Other works apply this idea to infer the detailed displacement maps~\cite{chen2019photo,cao2015real} or super-res the reflectance maps~\cite{huynh2018mesoscopic}.
GANtlitz~\cite{gruber2024gantlitz} train a StyleGAN-like~\cite{karras2020analyzing} generative model over the reflectance patches.
At test time, it can generate full-resolution reflectance maps.
In this paper, we train a diffusion model to model the high-quality facial reflectance prior at the patch level.
We propose a novel patch-level DPS technique to generate seamless full-resolution maps that best match the captured images using this patch level prior.

\section{Overview}
Following CoRA~\cite{han2024cora}, our method takes a single co-located smartphone and flashlight sequence captured around the face in a dim room as input.
Such a capture setup has the following advantages: \emph{i)} it is easy to use at home, \emph{ii)} the high-frequency smartphone flashlight provides rich cues for reflectance estimation~\cite{ramamoorthi2001signal}, and \emph{iii)} the captured images are shadow-free, which eases the inverse rendering process.
Given the captured images, we aim to reconstruct a high-quality relightable scan with a base mesh and a set of reflectance maps including diffuse albedo, specular albedo, and normal.
To this end, we train a diffusion model to model the distribution of high-quality facial reflectance at patch-level (Section~\ref{sec:diff_prior}) and use this model as a prior for facial appearance capture (Section~\ref{sec:recon}).

\section{Patch-Level Diffusion Prior}\label{sec:diff_prior}
Our goal is to train a diffusion model to model the distribution of high-quality facial reflectance maps.
Once trained, the diffusion model is fixed as a prior in the low-cost facial appearance capture process.
To this end, one can directly train a model on the Light Stage maps.
However, it is challenging in practice since the original maps are in ultra-high resolution like 4K by 4K.
In addition, the current Light Stage dataset only contains hundreds-level scans, which is too scarce to learn a generative model even with strong data augmentation~\cite{gruber2024gantlitz}.
To address these problems, we learn a diffusion model over patches randomly cropped from the original maps.
To compensate for global information loss caused by patches, we augment the diffusion model with UV coordinate maps as a condition.
In this way, we can train the diffusion prior at a significantly lower resolution like 256 by 256 with abundant training data, while keeping the local and global prior information.
We describe the details of our patch reflectance dataset in Section~\ref{sec:diff_prior:data} and the diffusion model training process in Section~\ref{sec:diff_prior:train}.

\subsection{Patch Reflectance Dataset}\label{sec:diff_prior:data}

\begin{figure}[t]
    \centering
    \includegraphics[width=0.475\textwidth]{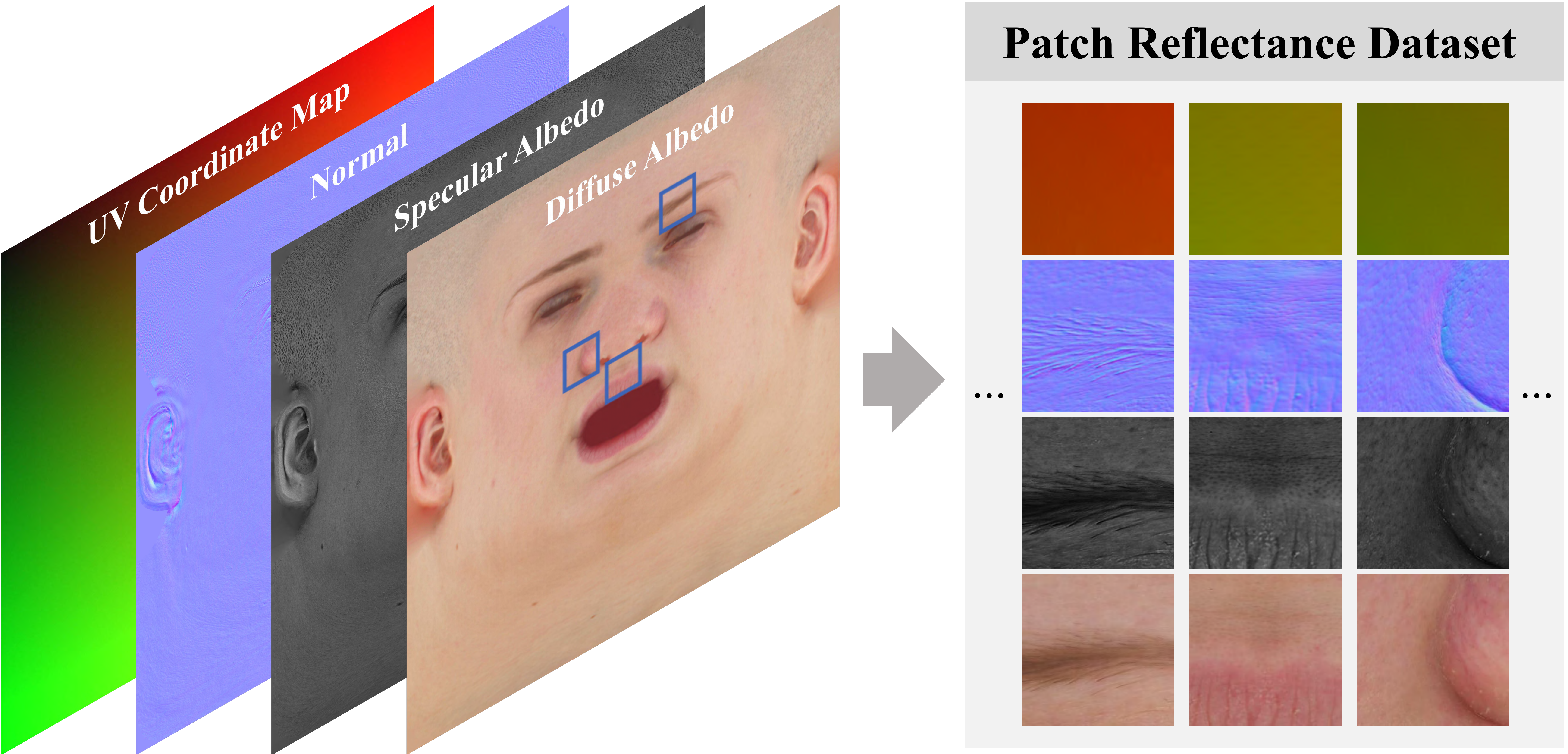}
    \caption{Overview of our patch reflectance dataset. It contains quadruples of diffuse albedo, specular albedo, normal, and the UV coordinate map randomly cropped from the full-resolution maps. }
    \label{Fig:data_preprocess}
\end{figure}

We purchase 48 Light Stage scans from 3D scanstore\footnote{https://www.3dscanstore.com/}, each with high-resolution diffuse albedo, specular albedo, and normal map.
We register all the scans to the ICT face topology~\cite{li2020learning} and resize the UV maps to 4K resolution.
The dataset contains 6 Asians (2 males and 4 females), 9 African Americans (5 males and 4 females), and 33 Caucasians (17 males and 16 females).
We outline the data processing steps in Figure~\ref{Fig:data_preprocess}.
For each subject, we concatenate all the reflectance maps and the UV coordinate map together along the channel dimension to form a 9-channel map, \emph{i.e.} 3 channels for diffuse albedo, 1 channel for specular albedo, 3 channels for normal, and 2 channels for UV coordinate. 
Then, we randomly crop 256 by 256 patches from these concatenated maps to form our patch reflectance dataset.
After processing, our dataset contains $10000$ quadruples of diffuse albedo, specular albedo, and normal patch along with the global position information recording where they cropped from, \emph{i.e.} the UV coordinate patch.

\subsection{Patch-Level Prior Training}\label{sec:diff_prior:train}
With the patch reflectance dataset, we adopt the diffusion model~\cite{ho2020denoising,Karras2022edm} to learn the prior considering its strong power in capturing data distribution and superior capability in solving inverse problems~\cite{daras2024survey}.
The diffusion model defines a forward diffusion process and a reverse diffusion process.
The forward diffusion process adds noise gradually to a data point $x_0$ sampled from the data distribution $q(x_0)$, which is described as a Gaussian transition:
\begin{equation}
\label{eq:diffusion_forward}
q\left(x_t \mid x_{t-1}\right):=\mathcal{N}\left(x_t ; \sqrt{1-\beta_t}\cdot x_{t-1}, \beta_t \cdot \mathbf{I}\right)
\end{equation}
Here, $\{\beta_t\in\left(0,1\right)\}_{t=1}^T$ are the fixed variance schedule.
With a large enough time step $T$, $q(x_T)$ is equivalent to the standard Gaussian distribution.
A nice property of the above process is that we can directly sample $x_t$ at any time step $t$ from $x_0$:
\begin{equation}
\label{eq:diff_forward}
q\left(x_t \mid x_0\right)=\mathcal{N}\left(x_t ; \sqrt{\bar{\alpha}_t}\cdot x_0,\left(1-\bar{\alpha}_t\right)\cdot\mathbf{I}\right)
\end{equation}
Here, $\alpha_t=1-\beta_t \text { and } \bar{\alpha}_t=\prod_{i=1}^t \alpha_i$.
In the reverse diffusion process, the diffusion model applies a neural network $\epsilon_\theta$ to learn the reverse of Eq.\eqref{eq:diffusion_forward}.
Specifically, it gradually denoises from the standard Gaussian noise $x_T$ to produce a clean data point $x_0$:
\begin{equation}
\label{eq:diff_reverse}
x_{t-1}=\frac{1}{\sqrt{\alpha_t}}\cdot\left(x_t-\frac{1-\alpha_t}{\sqrt{1-\bar{\alpha}_t}} \cdot\epsilon_\theta\left(x_t, t\right)\right)+\sigma_t\cdot z
\end{equation}
Here, $z$ is a standard Gaussian noise and the covariance $\sigma_t$ is predefined. 
In addition, at time step $t$, we can obtain the estimation of the clean data point $\hat{x}_t$ via:
\begin{equation}
\label{eq:diff_extimate_x0}
\hat{x}_t=\dfrac{1}{\sqrt{\bar{\alpha}_t}}\cdot(x_t-\sqrt{1-\bar{\alpha}_t}\cdot\epsilon_{\theta}(x_t,t))
\end{equation}

In our case, the data point $x_0$ is the concatenation of diffuse albedo, specular albedo, and normal patch along the channel dimension.
Recall that we apply a UV coordinate conditioned diffusion model to compensate for global information lost when split into patches, Eq.\eqref{eq:diff_reverse} and Eq.\eqref{eq:diff_extimate_x0} becomes:
\begin{align}
\label{eq:ours_diff_reverse}
x_{t-1}&=\frac{1}{\sqrt{\alpha_t}}\cdot\left(x_t-\frac{1-\alpha_t}{\sqrt{1-\bar{\alpha}_t}} \cdot\epsilon_\theta\left(x_t, t|y\right)\right)+\sigma_t\cdot z \\
\hat{x}_t&=\dfrac{1}{\sqrt{\bar{\alpha}_t}}\cdot(x_t-\sqrt{1-\bar{\alpha}_t}\cdot\epsilon_{\theta}(x_t,t|y))
\end{align}
Here, $y$ is the condition; in our case, it is the UV coordinate patch.
We adopt the EDM~\cite{Karras2022edm} architecture and directly train on the 256 by 256 resolution from scratch.
We modify their UNet denoiser in several aspects: \emph{i)} we set the number of output channels to 7 where 3 channels for the diffuse albedo, 1 channel for the specular albedo, and 3 channels for the normal, and \emph{ii}) we concatenate the positional-encoded~\cite{mildenhall2020nerf} UV coordinate patch to the UNet denoiser's input to support conditioning.

We show some generated patches on 3 pre-defined facial regions in Figure~\ref{Fig:ablat_gen}.
It demonstrates that the trained prior can generate high-quality reflectance patches indistinguishable from the training set.
In addition, our conditioning strategy can precisely control the position of the generated patch, which can largely constrain the search space and lead to better results when using this prior for the downstream task like reflectance estimation, as demonstrated in Figure~\ref{Fig:ablat_tiled_uv}.

\begin{figure*}[t]
    \centering
    \includegraphics[width=1.\textwidth]{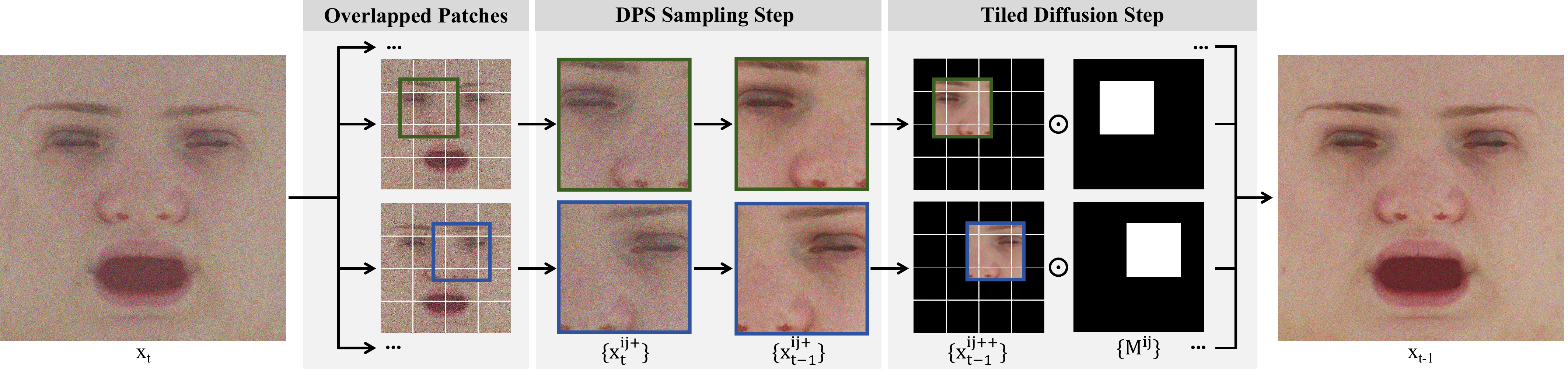}
    \caption{
        Pipeline of the proposed patch-level DPS technique. 
        At time step $t$, we split the full-resolution map $x_t$ into overlapped patches $\{x_t^{ij+}\}$; see Figure~\ref{Fig:split_patch} for a detailed illustration. 
        We then perform DPS to $\{x_t^{ij+}\}$ independently to obtain $\{x_{t-1}^{ij+}\}$.
        Next, we perform a Tiled Diffusion step to blend the overlapped regions in $\{x_{t-1}^{ij+}\}$ to obtain $x_{t-1}$. 
        Here we omit the condition $y$ and only visualize the diffuse albedo component in $x_t$ for clarity.
    }
    \label{Fig:patch_dps}
\end{figure*}

\begin{figure}[t]
    \centering
    \includegraphics[width=0.475\textwidth]{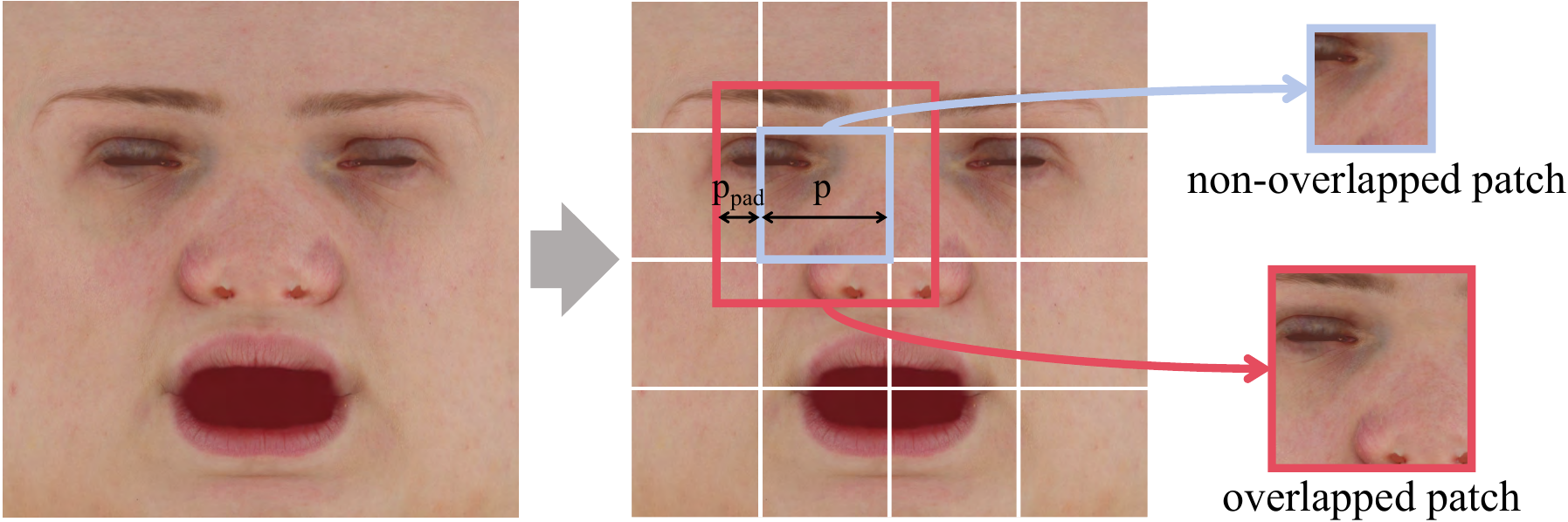}
    \caption{
    Illustration on overlapped-patch splitting.
    We first split the full map into non-overlapped patches with a size of $p$ by $p$.
    Then, we enlarge them along the border by $p_{pad}$ pixels to produce the overlapped patches. }
    \label{Fig:split_patch}
\end{figure}

\section{Prior-Guided Reconstruction}\label{sec:recon}
After training the diffusion model, we use it as a prior in the facial appearance capture process.
As our method takes the same input as CoRA, we first reconstruct facial geometry and scene lighting using CoRA (Section~\ref{sec:recon:geo_light}).
Then, we fix the geometry and lighting while only optimizing the facial reflectance maps within the distribution modeled by our diffusion prior (Section~\ref{sec:recon:ref}).

\subsection{Lighting and Geometry Reconstruction}\label{sec:recon:geo_light}
\subsubsection{Lighting Representation}
We use the same lighting model as CoRA.
Specifically, we separate the scene lighting into a high-frequency smartphone flashlight and a low-frequency ambient lighting.
The former is modeled by a point light source with predefined intensity $L$, and the latter is modeled by 2-order Spherical Harmonics (SH)~\cite{ramamoorthi2001efficient} in the $\rm SoftPlus$ output space to ensure non-negativity.
The observed color $l_o$ of the shading point $\rm \mathbf{x}\in\mathbb{R}^3$ is computed as:
\begin{align}
\label{eq:light_model}
    l_o &= l_{flash} + l_{amb} {\rm, where} \\
    l_{flash} &= \frac{L}{||\mathbf{x}-\mathbf{o}||_2^2}\cdot f_{pbr}(\mathbf{l},\mathbf{v},\mathbf{n},c,s,\rho)\cdot\max(\mathbf{n}\cdot\mathbf{v},0) \\
    l_{amb} &= c\cdot {\rm SoftPlus}(\sum_{l=0}^2\sum_{m=-l}^{l}K_{lm}\cdot Y_{lm}(\textbf{n}))
\end{align}
Here, $\mathbf{o}\in\mathbb{R}^3$ is the camera position, $\mathbf{v}\in\mathbb{R}^3$ is the view direction, $\mathbf{l}\in\mathbb{R}^3$ is the light direction\footnote{In the co-located setup, we have $\mathbf{l}=\mathbf{v}$ as the camera and the point light are in the same position.}, $\mathbf{n}\in\mathbb{R}^3$ is the normal direction, $c\in\mathbb{R}^3$, $s\in\mathbb{R}$, and $\rho\in\mathbb{R}$ are the diffuse albedo, specular albedo, and roughness of the Disney BRDF model $f_{pbr}$~\cite{burley2012physically},  $K_{lm}\in\mathbb{R}^3$ are the SH coefficients for the ambient light shading model, and $Y_{lm}(\cdot)$ are the SH bases.
We further split the above parameters into 3 groups.
The geometry parameters are $\mathbf{x}, \mathbf{o}, \mathbf{v}$ and $\mathbf{l}$.
The reflectance parameters are $c, s, \rho$ and $\mathbf{n}$.
The lighting parameters are $L$ and $K_{lm}$.

\subsubsection{Geometry Representation}
Considering that our high-quality reflectance prior is learned in a specific UV space, \emph{i.e.} the ICT face topology~\cite{li2020learning}, we adopt the triangle mesh with ICT's UV mapping as our geometry representation.

\subsubsection{Lighting and Geometry Reconstruction}
We first convert the captured video to $V$ frames $\{I_i\}_{i=1}^V$ and use MetaShape\footnote{https://www.agisoft.com/} to calibrate the camera parameters.
Then, we run CoRA on these frames to reconstruct geometry, reflectance map, and scene lighting.
We only use the reconstructed geometry and scene lighting while discarding the reflectance map because we will re-estimate it later within the distribution modeled by our diffusion prior.
Specifically, we directly use CoRA's lighting parameters since our lighting representation is identical to CoRA.
Considering that CoRA's geometry has inconsistent UV mapping with our prior model, as it is extracted from the neural SDF field~\cite{wang2021neus,yariv2021volume} using the Marching Cube algorithm~\cite{lorensen1998marching}, we register the ICT face template mesh~\cite{li2020learning} to it using Wrap\footnote{https://faceform.com/}.

\subsection{Reflectance Estimation with Patch-Level Prior}\label{sec:recon:ref}
With fixed facial geometry and scene lighting, we estimate the 4K resolution UV reflectance maps within the distribution modeled by the diffusion prior.
We optimize diffuse albedo $c$, specular albedo $s$, and normal $\mathbf{n}$ while using a fixed roughness map in our method as we do not have a prior for roughness $\rho$; we will discuss this design choice later in Section~\ref{sec:exp:lim}.
Formally, our goal is to guide the diffusion model to generate a reflectance map $x_0$ that best matches all the captured images $\{I_i\}_{i=1}^V$ after the rendering operation $\mathcal{R}$.
We implement $\mathcal{R}$ as a differentiable rasterization-based renderer~\cite{Laine2020diffrast} using the lighting model in Eq.\eqref{eq:light_model}.
With the geometry and lighting reconstructed from Section~\ref{sec:recon:geo_light}, given the view index $i$ and the reflectance map $x_0$, we can render it to an image $\hat{I_i}=\mathcal{R}(x_0,i)$; recall $x_0$ is the concatenation of the diffuse albedo, specular albedo, and normal map.

To this end, we adapt the diffusion posterior sampling (DPS) technique to our task.
Motivated by \citet{lyu2023diffusion}, we define the reverse sampling process of our diffusion model as:
\begin{align}
    x_{t-1}^{\prime} &= \frac{1}{\sqrt{\alpha_t}}\cdot\left(x_t-\frac{1-\alpha_t}{\sqrt{1-\bar{\alpha}_t}} \cdot\epsilon_\theta\left(x_t, t|y\right)\right)+\sigma_t\cdot z \label{eq:dps_sample} \\
    x_{t-1} &= x_{t-1}^{\prime} - \zeta_t \cdot\nabla_{x_t}\mathcal{L}_{pho}(\hat{x}_t) \label{eq:dps_grad}
\end{align}
Intuitively, it first denoises $x_t$ to a cleaner sample $x_{t-1}^{\prime}$ using Eq.\eqref{eq:ours_diff_reverse} and then moves it towards the direction such that the clean estimation $\hat{x_t}$ minimizes the photometric loss $\mathcal{L}_{pho}$ to some extent controlled by the step size $\zeta_t$.
Here, $\mathcal{L}_{pho}$ is defined as:
\begin{equation}
    \mathcal{L}_{pho}(x) = \sum_{i=1}^V||\mathcal{R}(x,i)-I_i||_2^2
\end{equation}
Although our patch-level diffusion prior can generate higher resolution results by sending higher resolution initial noise $x_T$ and UV coordinate map $y$ to it, directly applying Eq.\eqref{eq:dps_sample} and Eq.\eqref{eq:dps_grad} to sample the 4K reflectance map is infeasible in practice due to the memory limitation of the graphics card.
For example, we find an 80G NVIDIA A800 can process on no more than 1536 by 1536 resolution. 
To address this problem, we split the full-resolution
map into overlapped patches and propose a patch-based diffusion posterior sampling technique to sample them collaboratively.

\begin{figure*}[t]
    \centering
    \includegraphics[width=\textwidth]{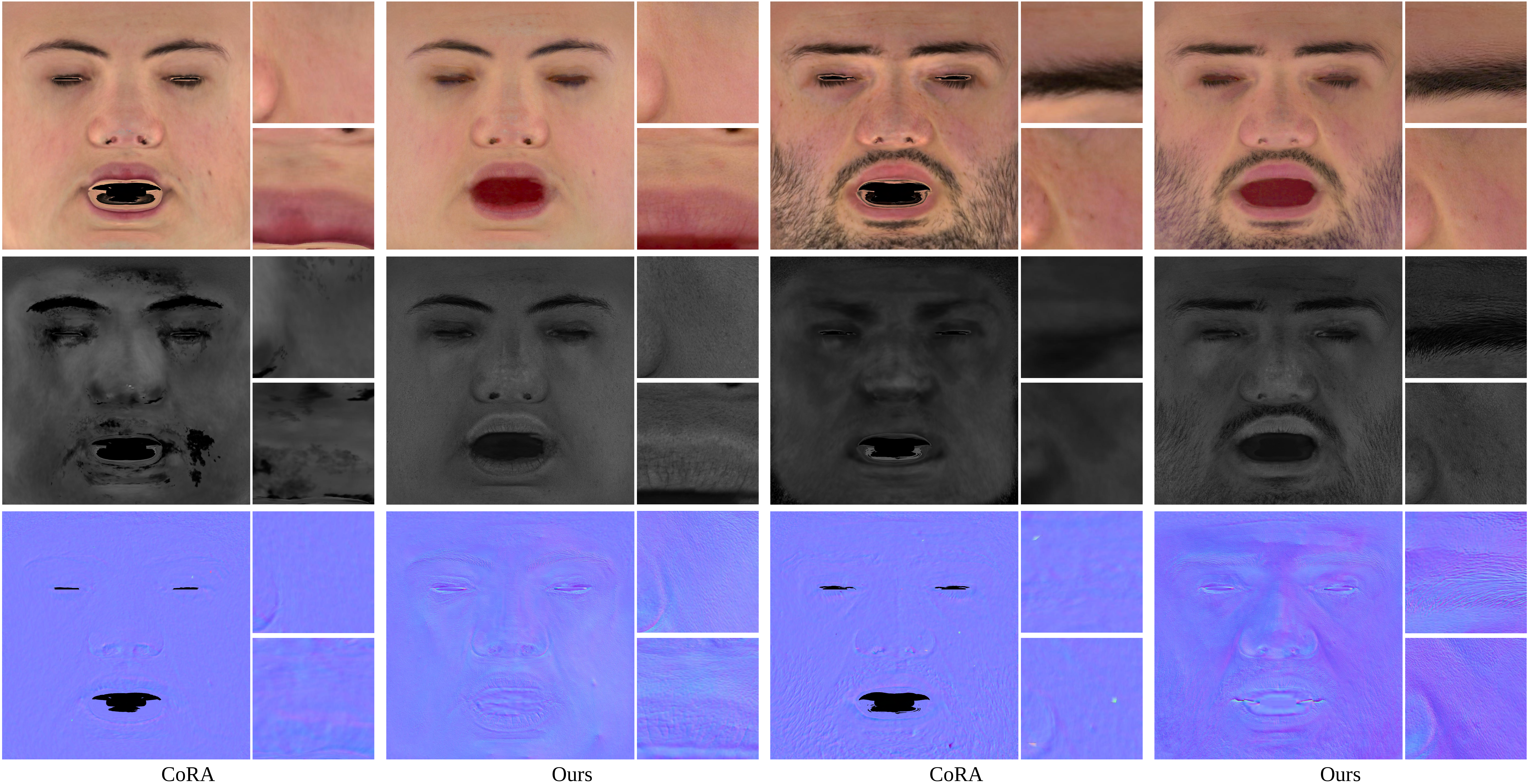}
    \caption{Qualitative comparison on reflectance map reconstruction of our method and CoRA~\cite{han2024cora}. From top to bottom: diffuse albedo, specular albedo, and normal.}
    \label{Fig:cmp_cora_map}
\end{figure*}

\subsubsection{Patch-Level Diffusion Posterior Sampling}
To reduce the memory cost, a naive solution is to split the full-resolution map $x_T$ and $y$ into $\frac{H}{p}\cdot\frac{W}{p}$ non-overlapped patches:
\begin{equation}
    \{x_T^{ij}\} \leftarrow x_T, \,\,\, \{y^{ij}\} \leftarrow y
\end{equation}
Here, $H=W=4K$ is the full map size and $p$ is the patch size.
Then, sample each patch independently to obtain $\{x_0^{ij}\}$ and fill them back to the full-resolution map $x_0$.
However, as shown in Figure~\ref{Fig:ablat_tiled_uv} (\emph{w/o overlap}), this leads to apparent seams in the results. 
That is because adjacent patches can easily converge to different local minima as solving reflectance parameters is ill-posed in our low-cost setup.

To address this problem, we draw inspiration from Tiled Diffusion~\cite{bar2023multidiffusion,jimenez2023mixture}, a training-free technique to generate high-resolution images from a diffusion model trained on low-resolution images.
In Tiled Diffusion, they split the high-resolution image into overlapped patches, and blend the overlapped region during the reverse diffusion process.
In this way, they can generate seamless high-resolution results.

We illustrate our method in Figure~\ref{Fig:patch_dps}.
Intuitively, we alternate between the DPS and the Tiled Diffusion step during sampling.
Specifically, at time step $t$, we first split the current reflectance map $x_t$ and condition $y$ into overlapped patches $\{x_t^{ij+}\}$ and $\{y^{ij+}\}$ via enlarging the original non-overlapped patches by $p_{pad}$ pixels; see Figure~\ref{Fig:split_patch} for detailed illustration.
Thus, the final patch size $p^+ = p + 2\cdot p_{pad}$.
Then, we perform a DPS step to $\{x_t^{ij+}\}$ independently according to Eq.\eqref{eq:dps_sample} and Eq.\eqref{eq:dps_grad} to obtain $\{x_{t-1}^{ij+}\}$.
Next, following Tiled Diffusion, we blend the overlapped regions in $\{x_{t-1}^{ij+}\}$ to obtain $x_{t-1}$.
Specifically, we pad $\{x_{t-1}^{ij+}\}$ to the full resolution to obtain $\{x_{t-1}^{ij++}\}$ and compute $x_{t-1}$ as the weighted average of $\{x_{t-1}^{ij++}\}$:
\begin{equation}
    x_{t-1} = \frac{\sum_{i,j} x_{t-1}^{ij++} \cdot M^{ij}}{\sum_{i,j}M^{ij}}
\end{equation}
Here, $M^{ij}$ is a binary mask that indicates the occupied region of the $ij$-th overlapped patch $x_{t}^{ij+}$ on the full-resolution map.
Thanks to our novel patch-level diffusion posterior technique, we can obtain seamless results as shown in Figure~\ref{Fig:ablat_tiled_uv} in a memory-efficient way.

\subsubsection{Efficient Implementation}
To enable DPS sampling at the patch level, our renderer $\mathcal{R}$ should support rendering a reflectance patch generated by the diffusion model, \emph{i.e.} $\hat{x_t^{ij+}}$, to screen space to compute the gradient of the photometric loss $\mathcal{L}_{pho}$.
A naive solution is to pad the reflectance patch to the full resolution and render it directly.
However, this is inefficient as we waste most of the computation on the padding region, which does not provide any gradients to the diffusion model.

To improve efficiency, our key observation is that to render the UV maps on the screen, the renderer $\mathcal{R}$ only needs to know the pixel-to-UV correspondence for each view.
Based on this, we first rasterize the UV coordinate map to the screen space to obtain the UV coordinate images corresponding to the full-resolution UV map.
This needs to be done only once as the geometry is fixed.
Given a generated reflectance patch, we online compute the translation and scale of the UV coordinate images such that sampling the reflectance patch with the transformed UV coordinate images gives the same results as sampling the full-resolution UV map with the original UV coordinate images.
With the UV position of the patch, the translation and scale have a closed-form solution.
This way, we can directly render the reflectance patch to the screen without padding.
In addition, we adopt mipmapping in $\mathcal{R}$ for texture sampling.

\begin{figure*}[t]
    \centering
    \includegraphics[width=\textwidth]{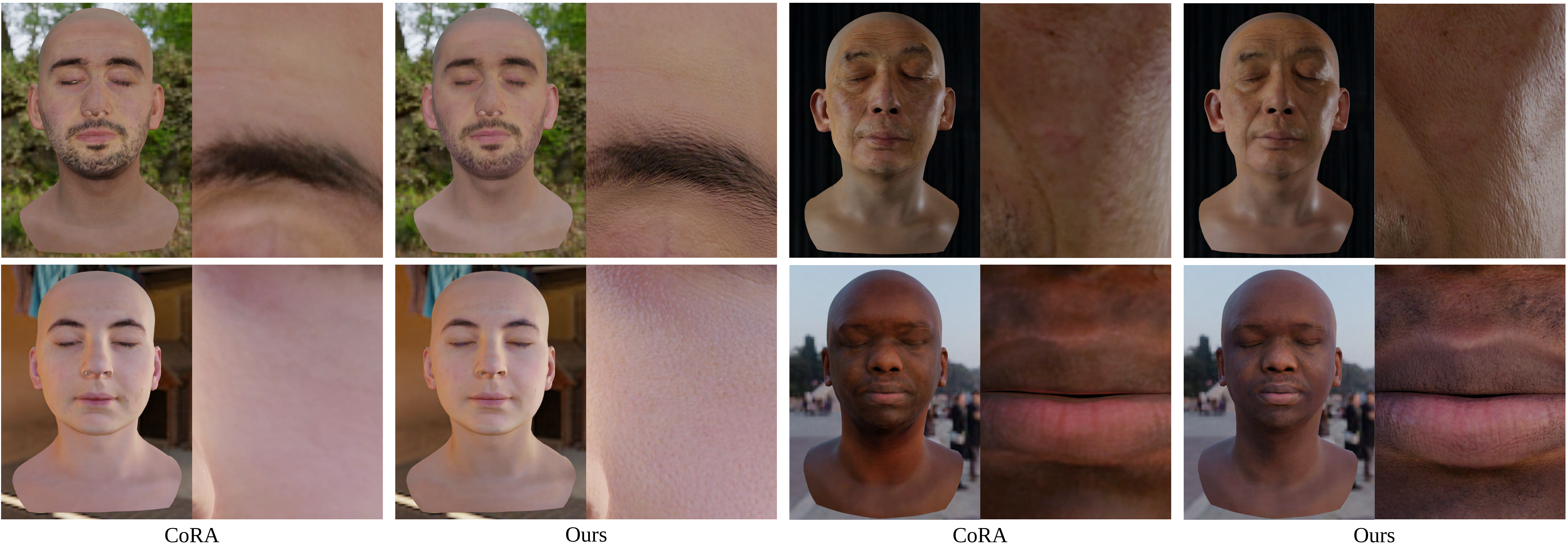}
    \caption{Qualitative comparison on face relighting of our method and CoRA~\cite{han2024cora}. }
    \label{Fig:cmp_cora_relight}
\end{figure*}

\section{Experiments}
In this section, we first introduce the implementation details of our method in Section~\ref{sec:exp:imp}.
Then, we compare our method with state-of-the-art methods with different settings in Section~\ref{sec:exp:cmp} and evaluate key design choices in our methods in Section~\ref{sec:exp:eval}.
Next, we test our method on diverse ethnic groups in Section~\ref{sec:exp:res} and present an application to create realistic full-head avatars in Section~\ref{sec:exp:app}.
Finally, we discuss our method's limitations and future works in Section~\ref{sec:exp:lim}.

\subsection{Implementation Details}\label{sec:exp:imp}
\subsubsection{Patch-Level Prior Training}
We adopt the K-Diffusion\footnote{https://github.com/crowsonkb/k-diffusion} implementation of EDM~\cite{Karras2022edm} and use its Image-v1 network architecture with the modifications mentioned in Section~\ref{sec:diff_prior:train}.
The training takes 24 hours to converge using 8 48G NVIDIA L20 with a total batch size of 96. 

\subsubsection{Prior-Guided Reconstruction}
We conduct all the experiments on a single 24G NVIDIA RTX 4090.
In the geometry and lighting reconstruction stage, we directly use the default config of CoRA~\cite{han2024cora} to process our data.
This stage takes about 60 minutes.
During reflectance estimation, we find 576 is the biggest patch size that can fit into the 24G memory.
We set the non-overlapped patch size $p$ to 448 and the padding size $p_{pad}$ to 64.
In this way, the actual patch size $p^+$ is 576.
We set the number of sampling steps $T=1000$.
Following \citet{chung2022diffusion}, we set the step size $\zeta_t$ to the reciprocal square root of $\mathcal{L}_{pho}$ at time step $t$.
We adopt the deterministic sampling strategy, \emph{i.e.} DDIM~\cite{song2020denoising}, to avoid inconsistency caused by random noise during sampling.
We uniformly sample $V=20$ views from the captured video for reflectance estimation.
Following CoRA, we resize all the images to 960 by 720 resolution.
This stage takes about 8 hours. 

\subsubsection{Dataset}
We contact the CoRA team~\cite{han2024cora} to obtain their dataset.
We also capture our own dataset using an iPhone X following the instructions provided by CoRA.
If not otherwise specified, the data is captured in a bedroom at night with curtains closed.
In total, our dataset contains 5 Asians (4 male and 1 female), 2 Caucasians (1 male and 1 female), and 1 African American (1 male).

\begin{figure}[t]
    \centering
    \includegraphics[width=0.475\textwidth]{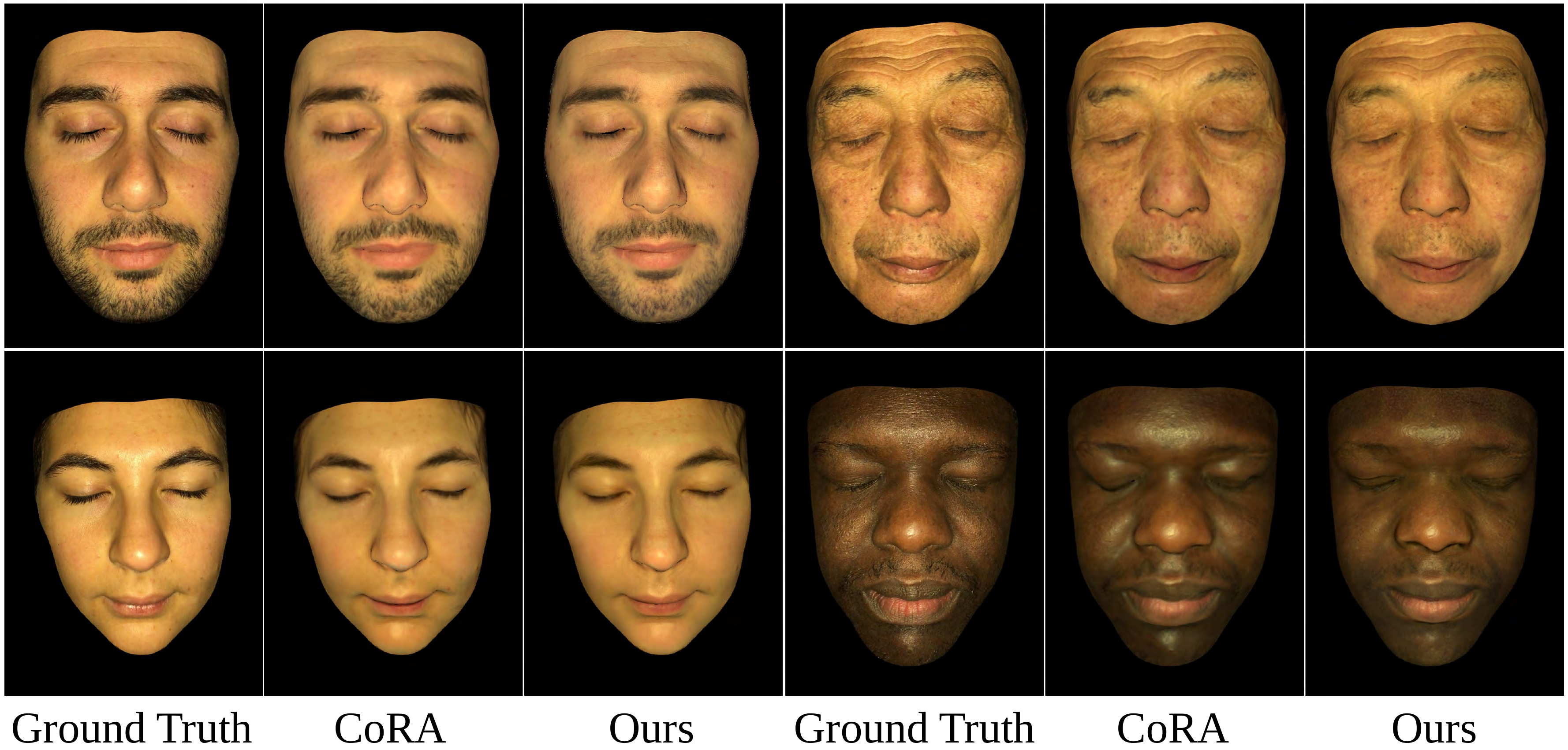}
    \caption{Qualitative comparison on face reconstruction of our method and CoRA~\cite{han2024cora}. }
    \label{Fig:cmp_cora_recon}
\end{figure}

\subsubsection{Metrics}
We report the SSIM, PSNR, and LPIPS~\cite{zhang2018perceptual} scores to measure the image quality.
We compute these metrics only on the facial skin region.

\subsection{Comparisons}\label{sec:exp:cmp}
We compare our method with various competitors, ranging from high-budget studio-based to low-cost single-view-based methods.
Firstly, we comprehensively compare our method to CoRA~\cite{han2024cora}, which is the state-of-the-art method with the same capture setup as ours.
We then compare our results with the Light Stage-based results~\cite{ghosh2011multiview} to evaluate the quality gap between ours and the high-end ones.
Lastly, we compare two more lightweight methods that can reconstruct facial geometry and appearance from in-the-wild images, \emph{i.e.} HiFi3Dface~\cite{hifi3dface2021tencentailab} and MoSAR~\cite{Dib_2024_CVPR}, to evaluate the performance gain between ours and these more easy-to-use ones.

\begin{table}[t]
\caption{Quantitative comparison on face reconstruction of our method and CoRA~\cite{han2024cora} over 8 captured subjects. Here, Cau. denote Caucasian; M and F denote male and female respectively. }
\begin{tabular}{cccclccc}
\hline
\multirow{2}{*}{Subject} & \multicolumn{3}{c}{CoRA}                               &  & \multicolumn{3}{c}{Ours}                               \\ \cline{2-4} \cline{6-8} 
                         & \!\!\!PSNR $\uparrow$\!\!\! & \!\!\!SSIM $\uparrow$\!\!\! & \!\!\!LPIPS $\downarrow$\!\!\! & & \!\!\!PSNR $\uparrow$\!\!\! & \!\!\!SSIM $\uparrow$\!\!\! & \!\!\!LPIPS $\downarrow$\!\!\! \\ \hline
Cau. M              & 32.52           & 0.9763          & 0.0552             &  & 33.28           & 0.9696          & 0.0589             \\
Cau. F              & 31.47           & 0.9468          & 0.0683             &  & 32.09           & 0.9380           & 0.0659             \\
Asian M1                 & 31.79           & 0.9541          & 0.0843             &  & 33.42           & 0.9439          & 0.0775             \\
Asian M2                 & 32.04           & 0.9585          & 0.0736             &  & 32.68           & 0.9533          & 0.0701             \\
Asian M3                 & 34.99           & 0.9733          & 0.0456             &  & 34.73           & 0.9687          & 0.0421             \\
Asian M4                 & 36.65           & 0.9817          & 0.0464             &  & 36.88           & 0.9763          & 0.0423             \\
Asian F                  & 32.02           & 0.9426          & 0.0705             &  & 32.87           & 0.9328          & 0.0723             \\
African M                  & 34.28           & 0.9434          & 0.0786             &  & 32.81           & 0.9315          & 0.0762             \\ \hline
Average                  & 33.22           & 0.9595          & 0.0653             &  & 33.59           & 0.9517          & 0.0631             \\ \hline
\end{tabular}
\label{Tab:cmp_cora}
\end{table}

\begin{figure*}[t]
    \centering
    \includegraphics[width=\textwidth]{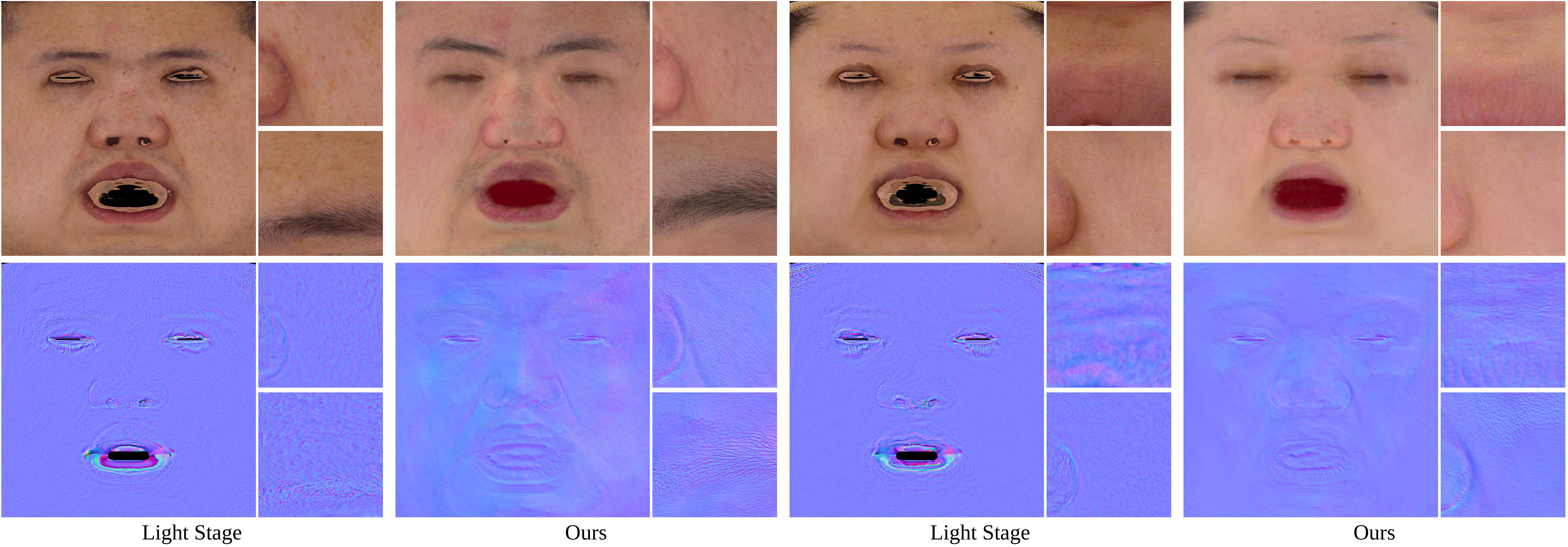}
    \caption{Qualitative comparison on the reconstructed diffuse albedo (top) and normal (bottom) of our method and a Light Stage method~\cite{ghosh2011multiview} implemented by \href{https://www.soulshell.cn/}{SoulShell.}}
    \label{Fig:cmp_light_stage}
\end{figure*}

\begin{figure*}[t]
    \centering
    \includegraphics[width=\textwidth]{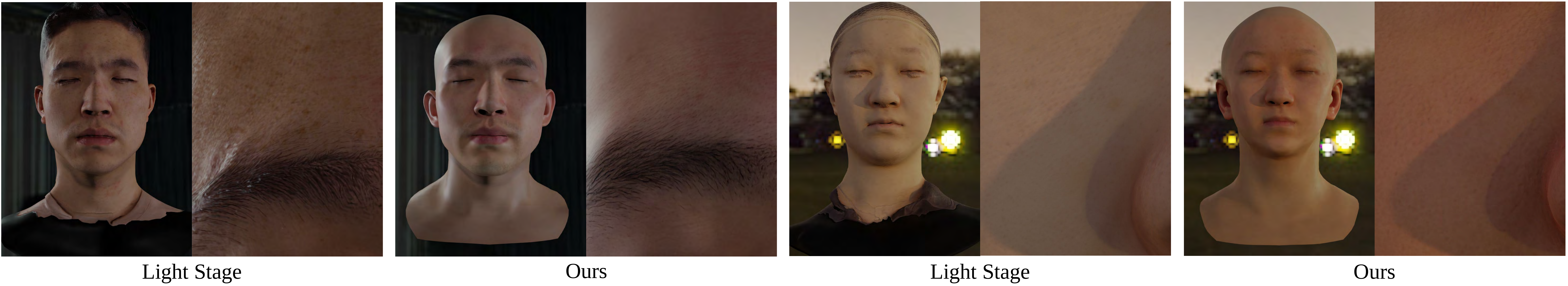}
    \caption{Qualitative comparison on face relighting of our method and a Light Stage method~\cite{ghosh2011multiview} implemented by \href{https://www.soulshell.cn/}{SoulShell.}}
    \label{Fig:cmp_light_stage_relight}
\end{figure*}

\subsubsection{Comparison to CoRA}
CoRA~\cite{han2024cora} takes the same input as our method.
It adopts a neural SDF field to represent facial geometry and a neural field to model the BRDF parameters as the reflectance.
To enhance diffuse-specular separation, the AlbedoMM~\cite{smith2020morphable} fitted specular albedo is introduced as a prior.
After training, it can export to a relightable scan with geometry and reflectance maps.
We compare the reconstructed reflectance maps of our method and CoRA in Figure~\ref{Fig:cmp_cora_map}; note we bake CoRA's results onto our mesh for better qualitative comparison.
CoRA can plausibly disentangle diffuse and specular components by using the AlbedoMM as a prior.
However, the reconstructed specular albedo is blurry due to the limited representation of the AlbedoMM prior, which is a PCA model.
In addition, it cannot produce high-quality diffuse and normal maps as it directly reconstructs them from the low-definition images captured by a smartphone video camera without using any prior.
Our method obtains significantly better results with more high-frequency details as we reconstruct reflectance maps within the distribution modeled by a high-quality diffusion prior.
As shown in Figure~\ref{Fig:cmp_cora_relight}, our method produces better results on face relighting as the reconstructed reflectance maps are of higher quality than CoRA; we conduct quantitative experiments on synthetic data in our \emph{supplementary material}.
In addition, we compare our method to CoRA on face reconstruction in Figure~\ref{Fig:cmp_cora_recon} and Table~\ref{Tab:cmp_cora} on the hold-out test views.
Although our method estimates reflectance maps in a more constrained space, it still obtains comparable results to CoRA.

\subsubsection{Comparison to Light Stage}
To evaluate the performance gap between our method and the state-of-the-art in the studio, we compare our method with the Light Stage method on two subjects.
Following CoRA~\cite{han2024cora}, we capture a co-located smartphone and flashlight sequence for the subject and run our method to obtain the results, and then scan the same subject using the Light Stage to obtain the corresponding Light Stage scan.
The Light Stage we used is an implementation of \citet{ghosh2011multiview} by \href{https://www.soulshell.cn/}{SoulShell}, where polarization filters are equipped in front of each light and camera to capture the facial reflectance.
Since their Light Stage cannot capture specular albedo, we only compare the reconstructed diffuse albedo and normal map.
Although the Light Stage we used is a degradation of \citet{ghosh2011multiview} as it cannot capture the specular albedo and specular normal map, it is the best one we have access to and we believe it can still provide insights on evaluating the quality of our method.
As shown in Figure~\ref{Fig:cmp_light_stage} and Figure~\ref{Fig:cmp_light_stage_relight}, our method demonstrates a slight quality gap to this Light Stage while significantly reducing the capture cost.

\begin{figure*}[t]
    \centering
    \includegraphics[width=\textwidth]{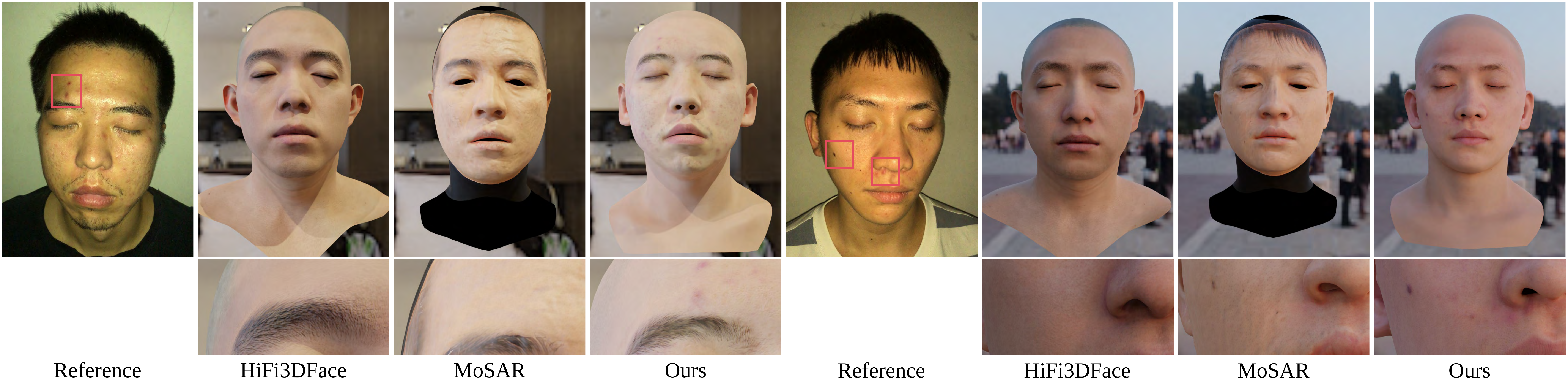}
    \caption{Qualitative comparison on face relighting of our method, HiFi3DFace~\cite{hifi3dface2021tencentailab}, and MoSAR~\cite{Dib_2024_CVPR}. }
    \label{Fig:cmp_hifi_mosar}
\end{figure*}

\begin{figure*}[t]
    \centering
    \includegraphics[width=1.\textwidth]{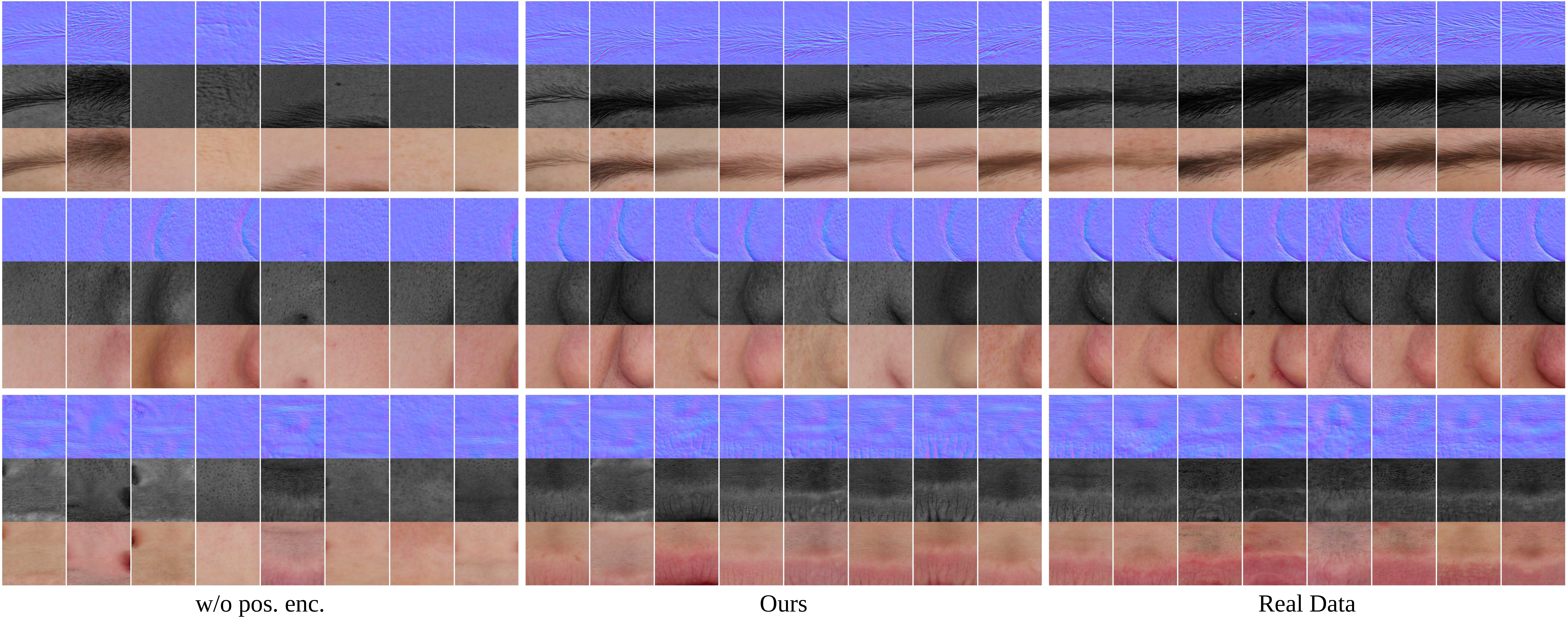}
    \caption{Qualitative ablation study of the network architecture on reflectance patch generation.
    For our method and the \emph{w/o pos. enc.} baseline, we generate 8 random samples at 3 pre-defined regions, \emph{i.e.} eyebrow, nose, and lip, shown from the top to the bottom row. 
    See their position on the full resolution map in Figure~\ref{Fig:data_preprocess}.
    We crop the same region from 8 randomly sampled reflectance maps in our Light Stage dataset as a comparison (the Real Data column).
    }
    \label{Fig:ablat_gen}
\end{figure*}

\subsubsection{Comparison to HiFi3DFace}
HiFi3DFace~\cite{hifi3dface2021tencentailab} reconstructs facial geometry and high-quality diffuse albedo and normal maps from multiple RGB images captured in the wild.
It trains a PCA model on a facial reflectance dataset to model the distribution of 2K by 2K diffuse albedo and normal maps and a Pix2Pix~\cite{isola2017image} model to add details to the PCA model's generation results.
During reconstruction, it first fits a parametric model to reconstruct the geometry.
Then, it optimizes the PCA model's parameter and applies the Pix2Pix model to enhance details.
We capture 4 extra images for each test subject with enough viewpoint coverage as the input to HiFi3DFace; we find that using more views does not improve their results.
We qualitatively compare our method to it on face relighting in Figure~\ref{Fig:cmp_hifi_mosar}.
Compared to HiFi3DFace, our method preserves more person-specific details like nevus and demonstrates higher-quality results.
That is because our diffusion prior has more expressive power than the PCA-based one used by their method.
In addition, the performance gain also comes from several system design choices in our method.
Firstly, we register a template mesh to the neural SDF as the geometry, which is more accurate than their parametric model-fitted one.
Secondly, our prior dataset is of higher quality and contains richer reflectance information than theirs; note our dataset contains 4K by 4K normal, diffuse albedo, and specular albedo maps, while their dataset only contains 2K by 2K diffuse albedo and normal maps.

\subsubsection{Comparison to MoSAR}
MoSAR~\cite{Dib_2024_CVPR} takes a single in-the-wild face image as input.
It first reconstructs the parametric geometry and then infers the reflectance maps from the unwrapped texture.
We capture an extra frontal view of each test subject and input it to MoSAR.
We qualitatively compare our method to it on face relighting in Figure~\ref{Fig:cmp_hifi_mosar}.
Compared to MoSAR, our method can reconstruct the captured subject more faithfully using significantly fewer training scans (we use 48 Light Scans for training while MoSAR uses 800+ scans).
That is because our multi-view setup provides more observations than the single-view setup in MoSAR, making our method significantly reduce the reliance on data-driven prior while better preserving the person-specific facial traits.

\begin{figure*}[t]
    \centering
    \includegraphics[width=\textwidth]{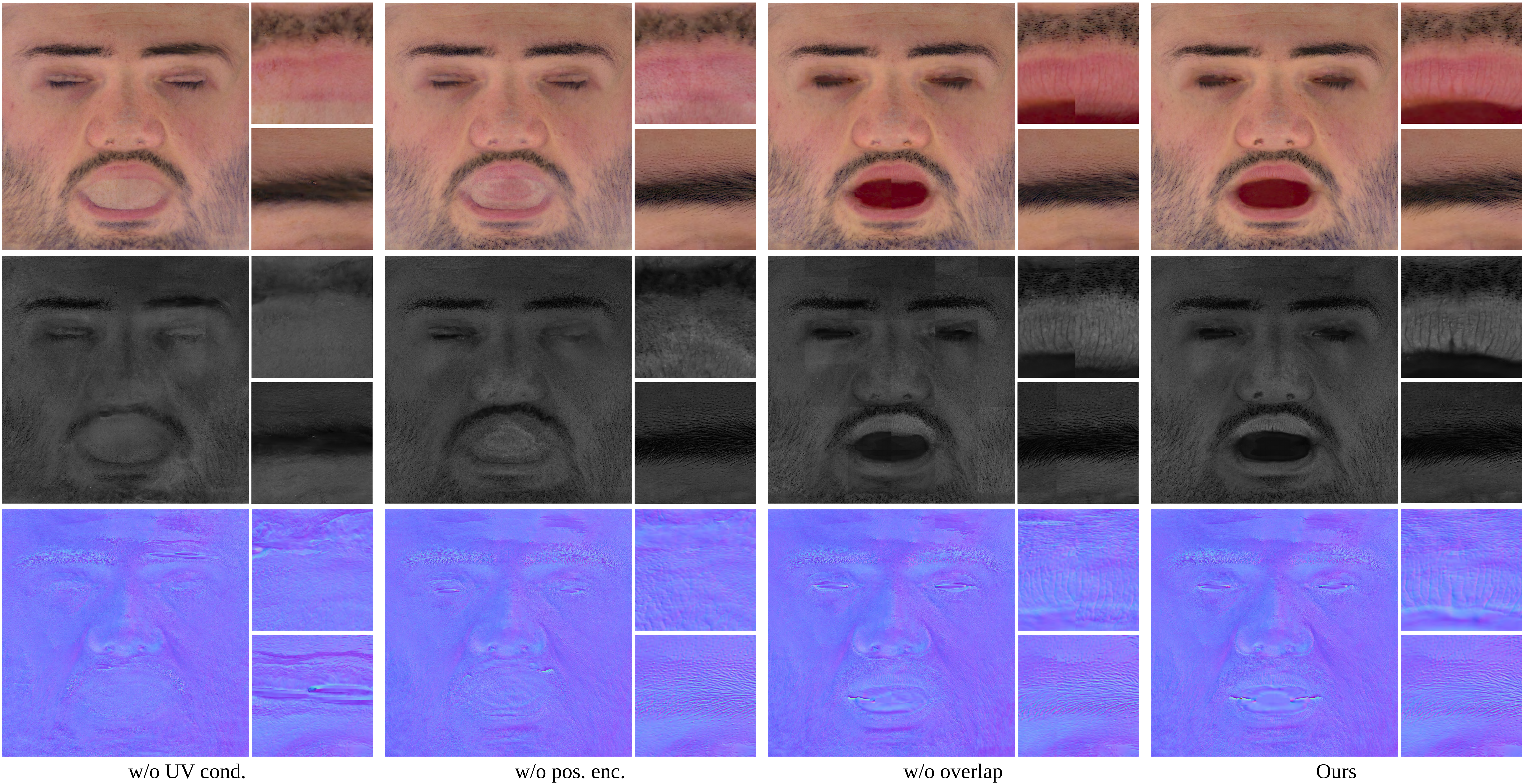}
    \caption{Qualitative ablation study of the key design choices in our method. }
    \label{Fig:ablat_tiled_uv}
\end{figure*}

\subsection{Evaluations}\label{sec:exp:eval}
In this section, we evaluate the core contributions of our method, \emph{i.e.} the design choice of introducing a UV-conditioned diffusion model as the prior and the patch-level DPS technique to steer this patch-level diffusion model to generate seamless full-resolution reflectance maps.
We first evaluate our diffusion prior, including the effectiveness of introducing the UV coordinate map as a condition and the choice of conditioning mechanism.
Then, we evaluate the proposed patch-level DPS technique, including the effectiveness of the Tiled Diffusion step in our method and the choice of hyperparameters. 

\begin{table}[t]
\caption{Quantitative ablation study of the design choices in our methods on face reconstruction. The metrics are averaged on 4 captured subjects.}
\begin{tabular}{ccccc}
\hline
Method & PSNR~$\uparrow$  & SSIM~$\uparrow$   & LPIPS~$\downarrow$  & Time (min)           \\ \hline
\emph{w/o UV cond.}       & 31.75 & 0.9541 & 0.071 & \multirow{2}{*}{508} \\
\emph{w/o pos. enc.}       & 31.61 & 0.9517 & 0.072  &                      \\ \hline
\emph{w/o overlap}       & 31.71 & 0.9530  & 0.071 & 508                  \\ \hline
$\zeta_t'=0.5$       & 30.32 & 0.9486 & 0.081 & \multirow{3}{*}{508} \\
$\zeta_t'=2$       & 32.74 & 0.9566 & 0.064 &                      \\
$\zeta_t'=5$       & 33.09 & 0.9526 & 0.065 &                      \\ \hline
$p=320$, $p_{pad}=128$       & 31.89 & 0.9549 & 0.069 & 938                  \\
$p=512$, $p_{pad}=32$       & 31.87 & 0.9531 & 0.071  & 433                  \\
$p=544$, $p_{pad}=16$       & 31.94 & 0.9533 & 0.070 & 391                  \\ \hline
Ours   & 31.87 & 0.9545 & 0.070 & 508                  \\ \hline
\end{tabular}
\label{Tab:ablat_quant}
\end{table}

\subsubsection{Evaluation on Diffusion Prior}
We first evaluate the design choice of introducing the UV coordinate map as a condition to our diffusion prior.
We introduce a baseline \emph{w/o UV cond.} where we train an unconditional diffusion model on our dataset and use it as a prior.
As shown in Figure~\ref{Fig:ablat_tiled_uv}, our method produces more details around the eyebrow and lip region on the reconstructed reflectance maps compared to this baseline.
Also, our method demonstrates better quantitative metrics in Table~\ref{Tab:ablat_quant}.
The reason is that solving high-quality reflectance parameters is ill-posed in our low-cost setup, as the captured images are low-definition and we do not have polarization filters for diffuse-specular separation.
With the UV coordinate map as a condition, our method searches the best-matched reflectance patch in a reduced space constrained by the UV coordinate map instead of the full space modeled by the diffusion prior. 
This makes the optimization easier and thus leads to better results.

We then evaluate how to introduce the UV coordinate map as a condition to the diffusion prior.
Because the UV coordinate map is spatially low-frequency, directly concatenating it to the denoiser's input does not work well; see \emph{w/o pos. enc.} in Figure~\ref{Fig:ablat_gen}. 
Using this naive conditioning mechanism leads to inferior reconstruction results as shown in Figure~\ref{Fig:ablat_tiled_uv} and Table~\ref{Tab:ablat_quant}.
Our method obtains superior results by making the condition more distinctive via positional encoding.

\subsubsection{Evaluation on Patch-Level DPS}
We first evaluate the effectiveness of the Tiled Diffusion step in the proposed patch-level DPS technique.
We design a baseline method \emph{w/o overlap}, where we directly split the full-resolution map into non-overlapped patches and sample each patch independently via DPS.
As shown in Figure~\ref{Fig:ablat_tiled_uv} and Table~\ref{Tab:ablat_quant}, this baseline method fails to produce seam-free results and obtains inferior metrics.
The reason is that adjacent patches can easily converge to different local minima as solving reflectance parameters is ill-posed in our low-cost setup. 
Our method significantly improves patch consistency as we collaboratively sample adjacent patches via the Tiled Diffusion step.
In addition, we compare to traditional methods for removing seams like Image Quilting~\cite{efros2023image} or Possion Editing~\cite{perez2023poisson} in our scenario in the \emph{supplementary material}.

\begin{figure}[t]
    \centering
    \includegraphics[width=0.475\textwidth]{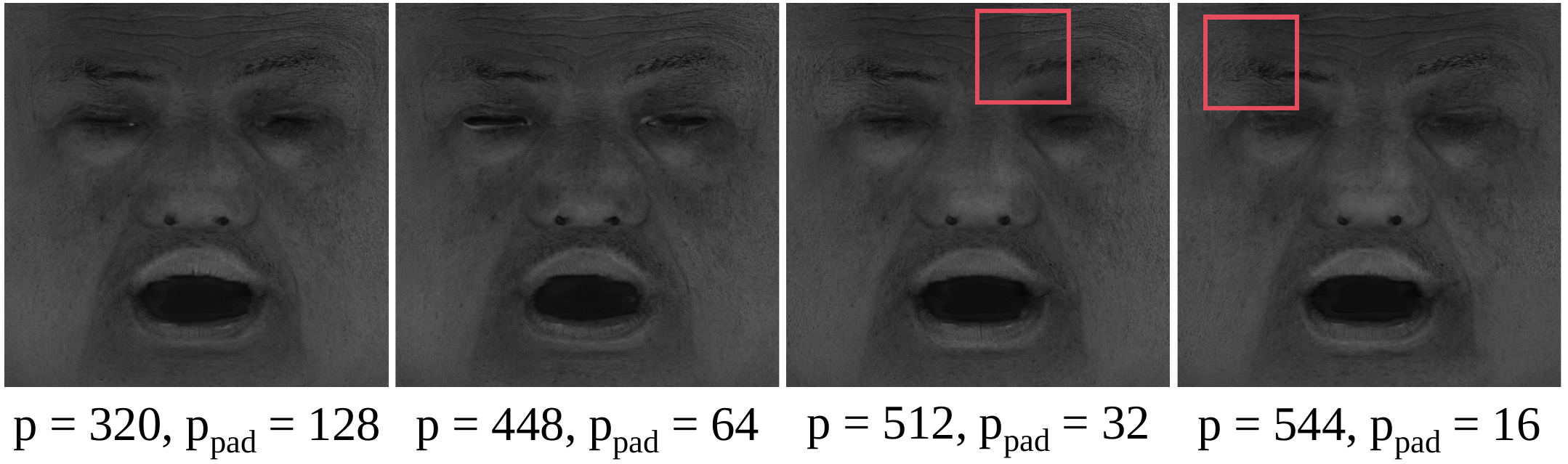}
    \caption{Qualitative ablation study of the patch size $p$ and the padding size $p_{pad}$. 
    We show the reconstructed specular albedo map here.
    }
    \label{Fig:ablat_patch}
\end{figure}

Then, we evaluate the choice of the patch size $p$ and the padding size $p_{pad}$.
We find the 24G NVIDIA RTX4090 can process on no more than 576 by 576 resolution.
So we test different combinations of $p$ and $p_{pad}$ such that the actual patch size $p^+$ is 576; recall $p^+=p+2\cdot p_{pad}$.
As shown in Table~\ref{Tab:ablat_quant}, different combinations of $p$ and $p_{pad}$ give similar photometric metrics.
However, the total runtime becomes shorter with bigger $p$.
That is because bigger $p$ leads to fewer patches and thus fewer repeated computations.
Thus, with the fixed actual patch size $p^+$, we prefer the combination of a bigger $p$ with a smaller $p_{pad}$ as it leads to a faster method.
However, as shown in Figure~\ref{Fig:ablat_patch}, small $p_{pad}$ leads to inferior quality as the overlapped region between adjacent patches is too small.
In practice, we choose $p=448$ and $p_{pad}=64$ to balance the efficiency and performance.

\begin{figure}[t]
    \centering
    \includegraphics[width=0.475\textwidth]{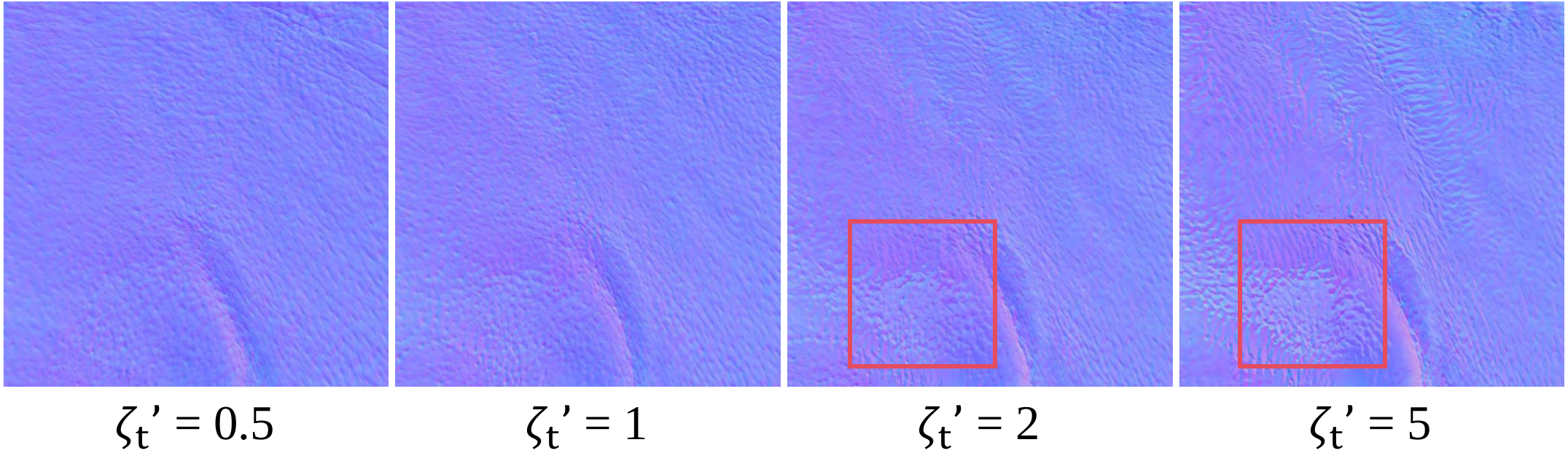}
    \caption{Qualitative ablation study of the guidance weight $\zeta_t'$. We show a close-up of the reconstructed normal map here. }
    \label{Fig:ablat_zeta}
\end{figure}

Lastly, we evaluate how the step size $\zeta_t$ affects the results.
$\zeta_t$ is used to control the guidance strength of the photometric loss.
Thus, we expect it can trade off the quality and fidelity.
We re-parameterize $\zeta_t=\frac{\zeta_t'}{\sqrt{\mathcal{L}_{pho}}}$ and compare the results obtained from different $\zeta_t'$.
As shown in Table~\ref{Tab:ablat_quant}, the quantitative metrics improve with bigger $\zeta_t$, which means the reconstructed maps can better explain the captured images.
On the other hand, smaller $\zeta_t$ leads to higher quality results as shown in Figure~\ref{Fig:ablat_zeta}.
We set $\zeta_t'=1$ to balance the quality and fidelity.

\subsection{More Results}\label{sec:exp:res}
\begin{figure*}[p!]
    \centering
    \includegraphics[width=0.9\textwidth]{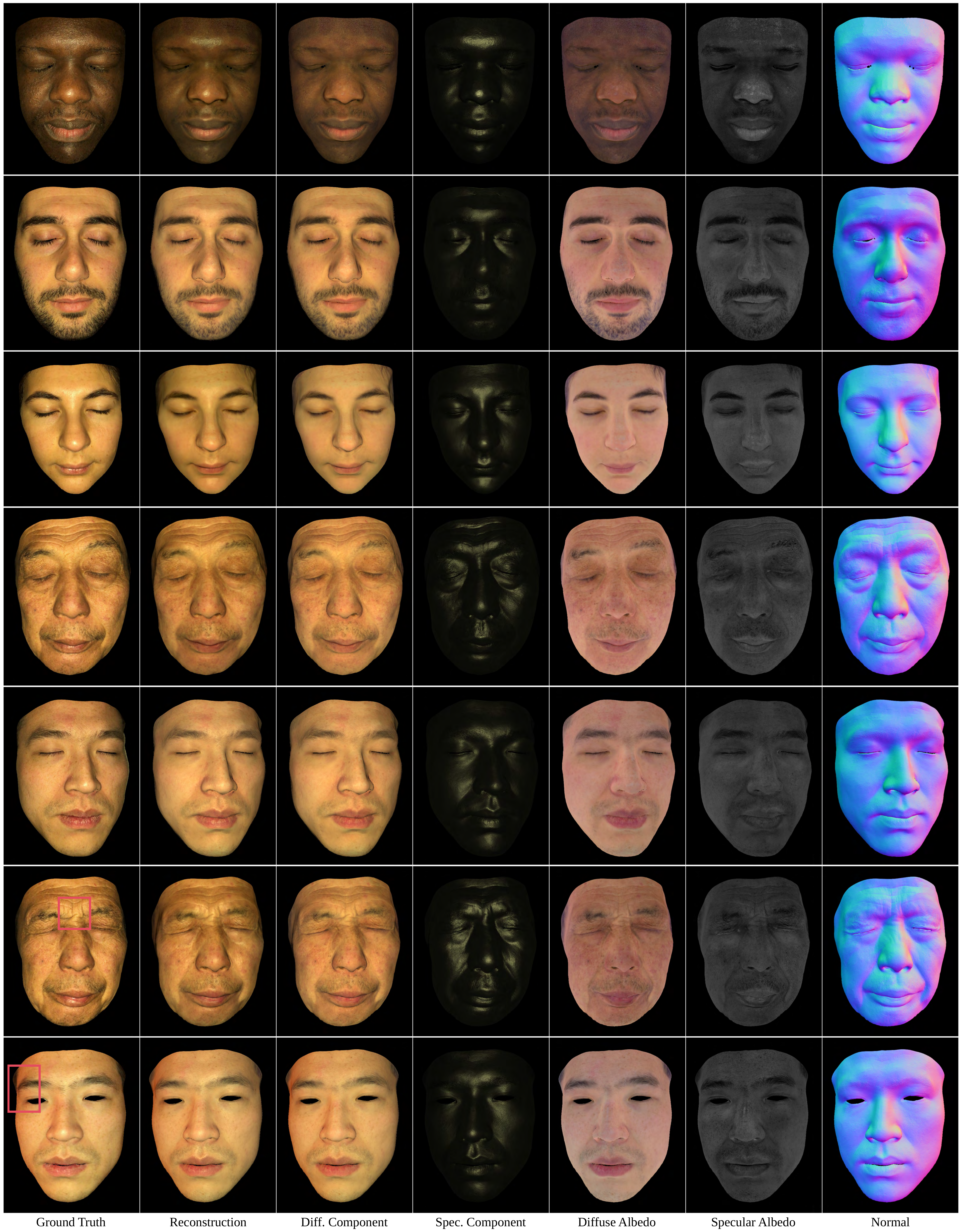}
    \caption{Results of our method on diverse ethnic groups. 
    Using the diffusion prior, our method can disentangle the diffuse and specular components from the captured RGB images in a plausible way, producing high-quality facial reflectance maps including diffuse albedo, specular albedo, and normal.
    }
    \label{Fig:ours_result}
\end{figure*}

We present the results of our method on diverse ethnic groups in Figure~\ref{Fig:ours_result}.
Although our diffusion prior is trained on only 48 Light Stage scans, it generalizes quite well to diverse ethnic groups.
In the 6th row of Figure~\ref{Fig:ours_result}, we test our method on a subject with eyes squeezing; see the 4th row as a reference of this subject captured with a neutral expression.
Although our training set only contains neutral scans, our method can even reconstruct facial expressions well.
We believe the following reasons explain why our method can obtain such good results using only 48 neutral training scans. 
Firstly, compared to previous works~\cite{Dib_2024_CVPR,lattas2020avatarme,lattas2021avatarme++} which learns a network to infer the reflectance maps from a single image, we explicitly solve the inverse rendering problem from the co-located smartphone and flashlight sequence, which makes our method significantly reduce reliance on data-driven prior as our problem is much more well-posed.
Secondly, the diffusion model can generate content beyond the training set~\cite{zhu2023unseen}; this property enables our diffusion model to generate person-specific facial traits identical to the observation to satisfy the photometric guidance, although they are unseen in the 48 training scans, making our results high-fidelity.

In addition, in the 7th row of Figure~\ref{Fig:ours_result}, we test our method on a subject captured under a challenging environment with asymmetrical lighting; see the 5th row as a reference of this subject captured under an easier environment.
Our method can still produce high-quality results in this case thanks to the explicit modeling of ambient illumination in our lighting model.
The right-side forehead looks red in the captured image, while our method successfully disentangles this effect from the ambient lighting and produces a clean result.

\begin{figure}[t]
    \centering
    \includegraphics[width=0.475\textwidth]{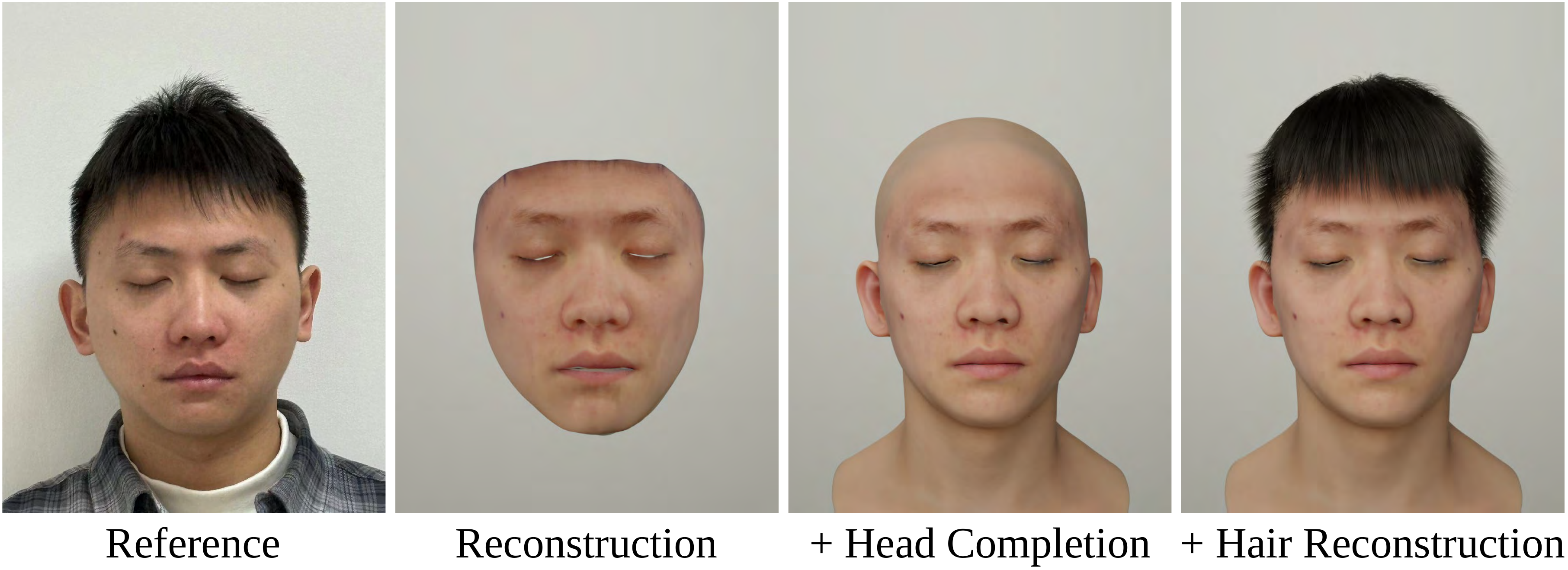}
    \caption{Application of our method. By combining our method with off-the-shelf head completion and hair reconstruction methods, we can create a realistic full-head avatar that is compatible with common graphics engines. }
    \label{Fig:hair}
\end{figure}

\subsection{Application: Realistic Full-Head Avatar}\label{sec:exp:app}
After reconstructing the high-quality facial scan using our method, we can obtain a photo-realistic full-head avatar with some post-processing steps.
We first complete the head geometry and texture maps and then run an off-the-shelf method to reconstruct the hair.
Below we give a detailed introduction to these steps; also see Figure~\ref{Fig:hair} for illustration.

\subsubsection{Head Completion}
For head geometry completion, we apply a two-stage method.
In the first stage, we fit a parametric model~\cite{zhang2023hack} to the reconstructed geometry.
In the second stage, we perform Laplacian deformation~\cite{sorkine2004laplacian} to the fitted parametric head such that the selected frontal facial-region vertices are close to the reconstructed geometry.
For head texture completion, we first train a PCA model on a full-head UV texture map dataset.
We then project the reconstructed reflectance maps to the PCA space to obtain the full-head maps.
Next, we blend the full-head maps with the reconstructed maps using a pre-defined frontal-facial region mask in the UV space.

\subsubsection{Hair Reconstruction}
We adopt Gaussian Haircut~\cite{zakharov2024gh} to reconstruct the hair.
Gaussian Haircut takes a monocular smartphone video captured around the subject's hair as input.
It can reconstruct a stand-based hair geometry from it.
In practice, we capture another video for the subject following the instructions provided by Gaussian Haircut.
We then run their method to obtain the reconstructed strand-based hair geometry.
We export it to Blender and manually align the hair geometry to the head.
For rendering, we manually tune the reflectance parameters of the hair.

\begin{figure}[t]
    \centering
    \includegraphics[width=0.475\textwidth]{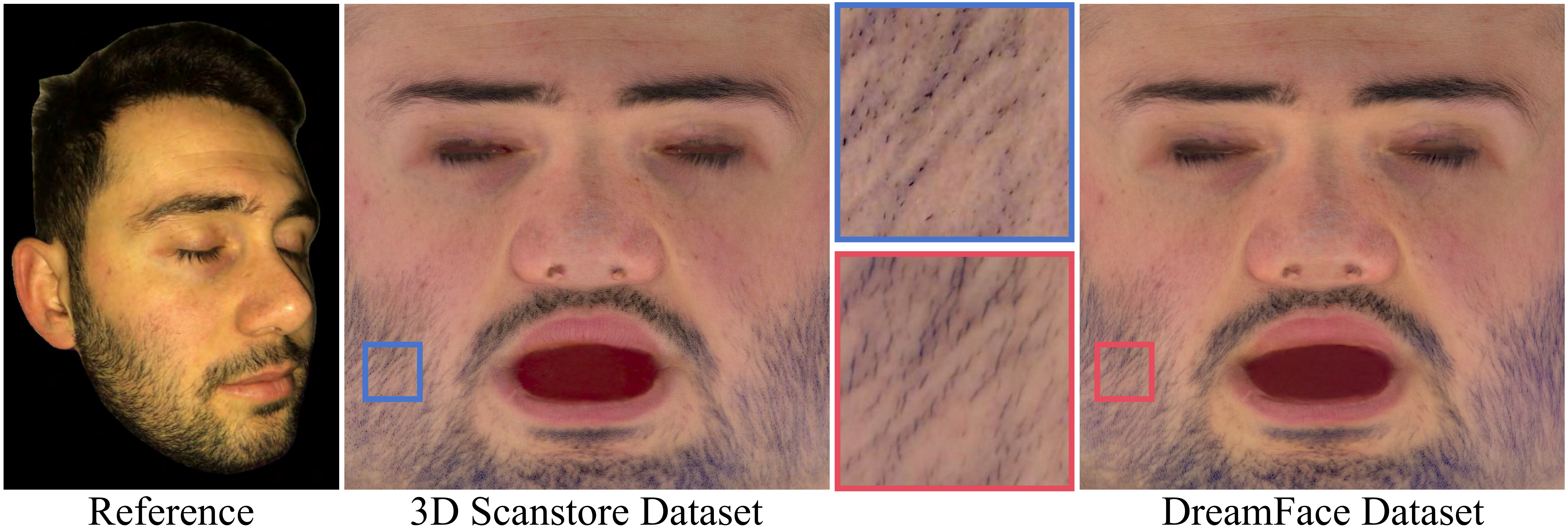}
    \caption{Limitation of our method. As the 3D Scanstore dataset does not contain big-beard subjects, the prior trained on this dataset cannot reconstruct the beard well. We try the DreamFace dataset to train the prior. The reconstruction quality improves around the beard region but the overall quality degrades due to the dataset quality difference.  }
    \label{Fig:beard}
\end{figure}

\subsection{Limitations, Discussions, and Future Works}\label{sec:exp:lim}

\paragraph{Runtime}
Our method currently takes about 8 hours to reconstruct the reflectance maps, which is time-consuming.
The reason is that DPS~\cite{chung2022diffusion} requires many steps to produce good results and the reflectance maps are of high resolution.
In the future, we can use more advanced sampling techniques like~\citet{rout2023secondorder} and adopt the coarse-to-fine strategy to solve low-resolution maps before going to the full resolution to reduce sampling steps.

\paragraph{Robustness to Misalignment in Data Capture}
As our capture setup relies on a single smartphone, the multi-view images are acquired at different timestamps, which may lead to misalignment due to the subtle movement of the subject.
In practice, extensive experiments on different subjects demonstrate that our method is robust to the subtle motion of an adult without special training when he or she tries to hold still during the data-capture period (typically 30 seconds).
We believe the reason is that the frames with unintentional subtle motion (\emph{e.g.} swallow and breathe) are much fewer than the frames where the subject is almost perfectly still; thus, the sampled $V=20$ training frames have a lower probability of containing these frames with subtle motion.
Using a motion detector to automatically clean the data is a possible future direction.

\paragraph{Geometry and Camera Calibration Error}
Although our method demonstrates high-quality results, we observe that it tends to bend the normal to compensate for the error in camera calibration and geometry reconstruction.
A possible future work is to jointly optimize camera parameters and facial geometry during the posterior sampling process of our diffusion prior.
We expect it can lead to a more visually pleasing tangent-space normal map.

\paragraph{Affects of the Data-Driven Prior}
As our method relies on data-driven prior, it cannot reconstruct high-quality content beyond the training set.
In practice, we find all the males from 3D scanstore have shaved the beard.
As shown in Figure~\ref{Fig:beard}, our method obtains inferior results on the bushy beard region using the diffusion prior trained on this dataset.
We also test our method on another Light Stage dataset.
We collect 100 DreamFace~\cite{zhang2023dreamface,qin2024instantfacialgaussianstranslator} generated scans and train another diffusion prior.
Using this prior, our method can better reconstruct the beard.
However, the overall quality is lower than the diffusion prior trained on 3D scanstore due to the dataset quality difference.
We believe a more diverse Light Stage dataset can solve the beard-reconstruction problem.

\paragraph{Fixed Roughness Map.}
We use a fixed roughness map as there is no prior for roughness in the Light Stage scan dataset we can access. We also experiment with optimizing a global roughness parameter like \citet{azinovic2023high} and \citet{riviere2020single}, but find the performance gain is marginal. We believe that solving a roughness map can further improve the results as demonstrated by previous works~\cite{weyrich2006analysis,han2023learning}; we leave it as our future work when a Light Stage scan dataset that contains roughness maps is available.

\paragraph{Global illumination and Subsurface Scattering.}
We currently adopt a local shading model and ignore the subsurface scattering effect in our method. 
As the captured smartphone and flashlight sequence are shadow-free, using a local shading model can still produce good results as we demonstrated in the paper.
We leave using a more advanced differentiable renderer~\cite{Mitsuba3} that supports simulating global illumination effects and subsurface scattering to further improve the results as our future work.

\section{Conclusions}
We propose a novel method for low-cost facial appearance capture.
Our key observation is that the primary reason low-cost methods produce lower-quality results than studio-based approaches is the lack of observation information.
To address this problem, we incorporate a data-driven prior trained on Light Stage scans to guide the reconstruction from low-cost captured data.
Specifically, we propose a diffusion model to model the distribution of high-quality reflectance maps at the patch level.
We condition this diffusion model with the UV coordinate map to compensate for the global information lost when splitting into patches.
Then, we propose a novel patch-level DPS technique to solve the reflectance map within the distribution modeled by the diffusion model.
We split the full-resolution map into overlapped patches and then alternate between the DPS step to minimize the photometric loss and the Tiled Diffusion step to maintain consistency.
Experimental results demonstrate our method obtains significantly better results than previous low-cost methods and largely close the quality gap to the high-budget methods in the studio.

\begin{acks}
    This work was supported by the National Key R\&D Program of China (2023YFC3305600), the Zhejiang Provincial Natural Science Foundation (LDT23F02024F02), and the NSFC (No.61822111, 62021002). This work was also supported by THUIBCS, Tsinghua University, and BLBCI, Beijing Municipal Education Commission. The authors would like to thank Jingwang Ling and Linjie Lyu for their insightful discussions and suggestions. The authors would also like to thank Wenbin Lin, Weixi Zheng, and Yudi Zhang for their extensive help on data capture. Feng Xu is the corresponding author.
\end{acks}

\bibliographystyle{ACM-Reference-Format}
\bibliography{sample-bibliography}

\end{document}